\documentclass[twocolumn]{IEEEtran}

\pdfoutput=1

\usepackage{amsmath,amsfonts,amsthm,amssymb}
\usepackage{mathtools}
\usepackage{bm}
\usepackage{mathrsfs}
\usepackage{graphicx}
\usepackage{epstopdf}
\usepackage{float}
\usepackage{caption}
\usepackage{balance}
\usepackage{subfig}
\usepackage{cite}
\usepackage{multirow}
\usepackage[ruled,linesnumbered]{algorithm2e}
\usepackage{enumerate}
\usepackage{xcolor}

\allowdisplaybreaks[4]

\usepackage{pdfcomment}
\hypersetup{hidelinks,
	colorlinks=true,
	allcolors=black,
	pdfstartview=Fit,
	breaklinks=true
}

\newcommand{\rev}{\textcolor{black}}

\begin{document}

\title{OTFS vs OFDM: Which is Superior in Multiuser LEO Satellite Communications}
\author{Yu Liu,~Ming Chen,~Cunhua Pan,~Tantao Gong,~Jinhong Yuan,~\IEEEmembership{Fellow, IEEE} and Jiangzhou Wang,~\IEEEmembership{Fellow, IEEE}
\thanks{ (Corresponding authors: Ming Chen, Cunhua Pan) }
\thanks{Y. Liu, M. Chen, C. Pan and T. Gong are with the National Mobile Communications Research Laboratory, Southeast University, Nanjing, 210096, China.
	(e-mail: \{liuyu\_1994, chenming, cpan, gongtantao \}@seu.edu.cn).}
\thanks{J. Yuan is with the School of Electrical Engineering and Telecommunications, University of New South Wales, Sydney, Australia. (j.yuan@unsw.edu.au).}
\thanks{J. Wang is with the School of Engineering, University of Kent, Canterbury, United Kingdom. (e-mail:j.z.wang@kent.ac.uk).}
}

\maketitle

\begin{abstract}
Orthogonal time frequency space (OTFS) modulation, a delay-Doppler (DD) domain communication scheme exhibiting strong robustness against the Doppler shifts, has the potentials to be employed in LEO satellite communications. 
However, the performance comparison with the orthogonal frequency division multiplexing (OFDM) modulation and the resource allocation scheme for multiuser OTFS-based LEO satellite communication system have rarely been investigated.
In this paper, we conduct a performance comparison under various channel conditions between the OTFS and OFDM modulations, encompassing evaluations of sum-rate and bit error ratio (BER). Additionally, we investigate the joint optimal allocation of power and delay-Doppler resource blocks aiming at maximizing sum-rate for multiuser downlink OTFS-based LEO satellite communication systems.
Unlike the conventional modulations relaying on complex input-output relations within the Time-Frequency (TF) domain, the OTFS modulation exploits both time and frequency diversities, i.e., delay and Doppler shifts remain constant during a OTFS frame, which facilitates a DD domain input-output simple relation for our investigation.
We transform the resulting non-convex and combinatorial optimization problem into an equivalent difference of convex problem by decoupling the conditional constraints, and solve the transformed problem via penalty convex-concave procedure algorithm. 
Simulation results demonstrate that the OTFS modulation is robust to carrier frequency offsets (CFO) caused by high-mobility of LEO satellites, and has superior performance to the OFDM modulation. Moreover, numerical results indicate that our proposed resource allocation scheme has higher sum-rate than existed schemes for the OTFS modulation, such as delay divided multiple access and Doppler divided multiple access, especially in the high signal-to-noise ratio (SNR) regime.
\end{abstract}


\section{Introduction}
Satellite communication technology has \rev{received extensive research attention}, which is considered to expand terrestrial communication networks into integrated space-ground communication networks. 
\rev{Due to its advantages of wide coverage, seamless connectivity and robustness for different geographical conditions, satellite communication has become a promising technology in the beyond fifth generation (B5G) and sixth generation (6G) communication systems to support the exponentially increasing data rate requirement and variety of users. }
Since the satellite communication system has been \rev{successfully} applied in mass broadcasting, navigation, and disaster relief operation \cite{SatCom1,SatCom2,SatCom3}, some researchers \rev{started to} incorporate it into B5G and future networks, a substantial number of standardization works on satellite communications are currently underway \cite{SatCom4,Satcom5}. 

\rev{According} to the orbit altitude, satellites are classified into three categories, low-earth-orbit (LEO) satellites ($500 \sim 2500$ km), medium-earth-orbit (MEO) satellites ($5000 \sim 20000$ km) and geostationary-earth-orbit (GEO) satellites (35786 km) \cite{HanZhuSatComLiteratureReview}. Compared to traditional GEO satellites and MEO satellites, \rev{the emerging LEO satellites provide the devices access with low propagation delay, low path loss, and flexible elevation angle \cite{Di2019UltraDenseLEO}.} These advantages \rev{enable} the LEO satellites to be a non-trivial infrastructure in constructing future integrated space-terrestrial networks \cite{SAGIN1,SAGIN2}.
Nowadays, there have been \rev{numerous} companies planning to launch satellites, such as spaceX and oneWeb \cite{spaceX,oneWeb}. 
However, unlike GEO satellites, LEO satellites circle at the earth's orbit with a velocity of 7.8-8 km/s, which is less than the speed of earth's rotation. Consequently, LEO satellites do not appear stationary with respect to the ground users, and this relative motion between LEO satellites and the ground users leads to significant Doppler shifts. According to the 3GPP standard TR 38.811, the Doppler shifts can reach up to $\pm 48$ kHz at the S band, and can increase to $\pm 480$ kHz at the Ka band \cite{3GPPTR38.811}. 
The Doppler shifts can result in the carrier frequency offsets (CFO), which may induce cyclic shifts and inter-carrier interference (ICI). \rev{These effects eventually lead to the significant degradation of the transmission performance.}
Henceforth, due to the severe Doppler shifts, the orthogonal frequency division multiplexing (OFDM) modulation \cite{2012Wang1,2012Wang2} does not seem to be an appropriate choice, and suitable alternative should be considered. The orthogonal time frequency space (OTFS) modulation has been recently proposed as a new modulation scheme specifically designed to mitigate the effect of Doppler shifts in the linear time-varying (LTV) channel \cite{OTFSsurvey1,OTFSsurvey2}.

OTFS was proposed by Hadani et al. in \cite{2017OTFSFundamentals}, which exploited full diversity over time and frequency to convert the time-varying fading channels experienced by modulated signals into a time-independent channel with a complex channel gain that is roughly constant for all symbols. This design allows the OTFS modulation to significantly improve the performance, particularly in systems with high Doppler shifts. 
In the recent years, the OTFS modulation has been further explored and extended to \rev{various} fields. In \cite{2018FarhangOTFSModem}, the authors \rev{presented} a discrete-time formulation of the OTFS modulation, which leads to a simplified transmitter. 
In \cite{2018Iterativedetection,2018InterferenceCancellationandIterativeDetection}, the authors explicitly \rev{derived} the input-output relations for reduced zero padding (RZP)/ reduced cyclic prefix (RCP) OTFS. Then in \cite{input-outputRelation}, the authors \rev{extended} the input-output relation of RZP/RCP OTFS to the matrix form. 
Meanwhile, the authors of \cite{2018RezazadehreyhaniOTFSmodem} \rev{derived} the relation for ZP/CP OTFS.
Furthermore, the authors of \cite{2021MohammedZak,2022LampelZak} \rev{presented} the precise relation and analysis of the OTFS modulation associated with the Zak transform \cite{ZakTransform}. The Zak transform enables the information symbols to interact with the delay-Doppler (DD) domain channel responses directly, which leads to a much simpler input-output relation over the complex high mobility channels compared to that of the OFDM modulation. More \rev{importantly}, the OTFS modulation enables the symbols principally \rev{to experience} the whole fluctuations of the channels in the time-frequency (TF) domain, which \rev{demonstrated} that the OTFS modulation offers the potentials of achieving full TF diversity \cite{2019SurabhiTFdiversity,2020HongTFDiversity,2021LiTFDiversity}.

\rev{Benefited from the appealing properties} of channels in the DD domain, including compactness, sparsity, quasi-stationary and separability, simple channel estimation and reduced-complexity detection approaches can be implemented. Initially, the authors of \cite{2019HongEmbeddedPilot} \rev{proposed} to embed the pilots into the resource grid in the DD domain to estimate the channel, which requires a sufficiently large guard interval around pilots to avoid the interference between pilots and information symbols. In \cite{2021WangChannelEstimation}, the authors \rev{applied this scheme to} the massive multiple-input and multiple-output (MIMO) OTFS system and \rev{proposed} a modified sensing matrix based channel estimation algorithm to acquire channel state information. To improve the transmission efficiency, the authors of \cite{2021YuanSuperimposed} \rev{proposed} to replace the embedded pilot by the superimposed pilot. In \cite{2022YuanOffgrid}, the authors \rev{considered} both the on-grid and the off-grid delay and Doppler shift components to perform sparse signal recovery with a sparse Bayesian-learning-assisted estimator. \rev{With respect to symbols detection, the message passing algorithm (MPA) and its variants, which model the characteristics of the interference in the DD domain via the Gaussian approximation, have been widely employed \cite{2018InterferenceCancellationandIterativeDetection,2021LiHybridMAP,2022YuanUnitaryAMP}.} These algorithms \rev{exploit} the channel sparsity in the DD domain \rev{such} that a \rev{satisfactory} error perfotrmance can be achieved with lower complexity. In addition, the authors of \cite{2020HongMRC} \rev{applied the} maximum ratio combining (MRC), a linear diversity combining technique, to symbols detection \rev{to reduce} the complexity. Additionally, the authors of \cite{2020SurabhiLMMSE} \rev{proposed} another linear minimum mean square error (LMMSE) algorithm with reduced complexity. 

In the recent years, a few \rev{contributions employed} the OTFS modulation \rev{to} the LEO satellite communications. The authors \rev{of} \cite{2023ZhouGrantFree} \rev{studied} the grant-free access for the OTFS-based LEO satellite communication system. In \cite{2022HongSatCom}, the authors \rev{proposed to employ the whitening transformation} at the transmitter and the receiver to combat the spatial correlations.  
In the term of multiple satellites system serving the single user, the authors of \cite{2023GCSatelliteDiversity} \rev{compared} the link performance of the OTFS modulation to that of the OFDM modulation, which indicates that link performance benefits from the OTFS modulation and satellites diversity. 
However, the comparison in this article is still not comprehensive. Primarily, it fails to take into account impact of diverse LTV channel conditions, notably neglecting the variations in CFO induced by high-velocity satellites movements. Moreover, \rev{this work did} not adequately investigate the influence of assorted post-processing techniques on the efficiency of communication performance, thereby restricting the scope of its findings. 

Despite ample research of channel estimation and symbols detection, there still lacks investigations on resource allocation with the OTFS modulation. 
Traditional resource allocation algorithm will result in excessive interference, which may lead to significantly degradation of transmission rates in practice.
Unfortunately, the design of resource allocation algorithm for the OTFS modulation is challenging. This is due to \rev{the fact} that the OTFS modulation does not guarantee the interference-free transmission like the OFDM modulation in static channels. Indeed, in the DD domain communications, received symbols generally contain the interference caused by multiple paths, as the result of the \rev{``twisted convolution''} operation \cite{2017OTFSFundamentals}. Consequently, the designs of resource allocation allocation algorithm for the OTFS modulation will face a complicated input-out relation and an effective channel matrix with a huge size, e.g., a number of delay bins times the number of Doppler bins rank square matrix.
The authors of \cite{2019OTFSMAMohammed,2022OTFSNOMARA} \rev{attempted} to circumvent this issue through a novel multiple access scheme, but it is only effective in the uplink transmission. 
In this paper, we consider the performance analysis and resource management algorithm design for a downlink multiuser (MU) LEO satellite communication system to address the above issues.
The major contributions of this paper can be summarized as follows:
\begin{itemize}
	\item We derive a succinct input-output relation for the downlink MU-OTFS LEO satellite communication system in matrix form and extend it to a symbol-wise format. 
	This relation forms the foundations for the performance analysis, and is also utilized for our resource allocation and symbols scheduling algorithm design.
	\item Based on the derived relation, we conduct a detailed interference pattern analysis within the DD domain. In \rev{particular}, we classify the interference suffered by the received symbols into two types, namely multipath self-interference (MPSI) and multiuser-interference (MUI), and reveal the physical meanings of them. For comparative purposes, we also derive the interference pattern for the OFDM modulation in the TF domain.
	\item Utilizing representative theoretical-equivalent models, we derive an expression for the achievable sum-rate for both the OTFS and OFDM modulations. \rev{Moreover, we conduct simulations to assess the system's achievable sum-rate and realistic bit error rate (BER), which indicates that the transmission performance using the OTFS modulation better than that of the OFDM modulation.} Additionally, the OTFS modulation demonstrates robustness in a variety of channel conditions, in contrast to the OFDM modulation. 
	\item We propose a penalty convex-concave procedure based algorithm for power allocation and symbols scheduling in the OTFS modulation. Simulation results demonstrate that the proposed algorithm can further enhance the sum-rate and BER performance for the OTFS system, especially in challenging channel conditions.
\end{itemize}

\textit{Notations}: The blackboard bold letters $\mathbb{A}$, $\mathbb{C}$ and $\mathbb{E}$ represent the constellation set, the complex number filed and the expectation operator respectively; the notation $\left(\cdot\right)^{\dagger}$ denotes the Hermitian transpose for a matrix; $\mathrm{vec}(\cdot)$ denotes the vectorization operator; $\mathrm{vec}_{M,N}^{-1}\left(\cdot\right)$ denotes the operation converting a $MN$ length vector to a $M\times N$ matrix; $\mathrm{sgn}(\cdot)$ represents the indicator function; $\mathrm{diag}\{\cdot\}$ denotes the diagonal matrix; $\otimes$ denotes the Kronecker product; $\mathrm{max}(\cdot)$ denotes maximization operator; $[\cdot]_M$ denotes the modulo operation with respect to $M$; $\lfloor\cdot\rfloor$ denotes the round down operation; $\delta(\cdot)$ is the Dirac function.

\section{System Model}
In this section, we derive a concise system model for MU-OTFS transmissions utilizing single-input and single-output (SISO). \rev{To this end, we firstly introduce the system structure and the LEO-satellite communication channel model, which will then be used for the related discussions on MU-SISO-OTFS transmissions.}

\subsection{System Structure}

\begin{figure}[hpbt]
	\centering
	\includegraphics[width=0.65\linewidth]{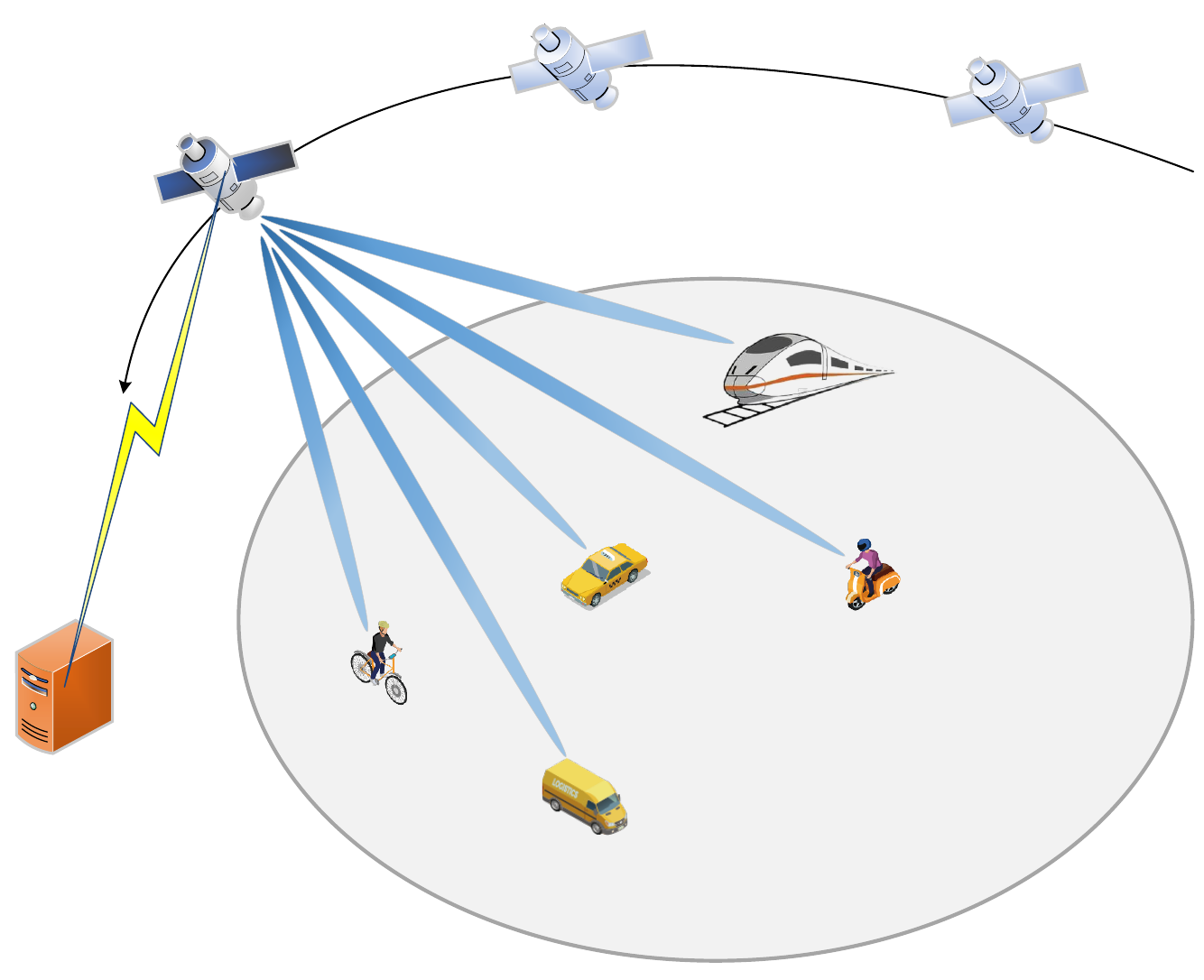}
	\caption{System Diagram of the Satellite-Terrestrial Network.}
	\label{diagram}
\end{figure}

\begin{figure*}[hpbt]
	\centering
	\includegraphics[width=1\linewidth]{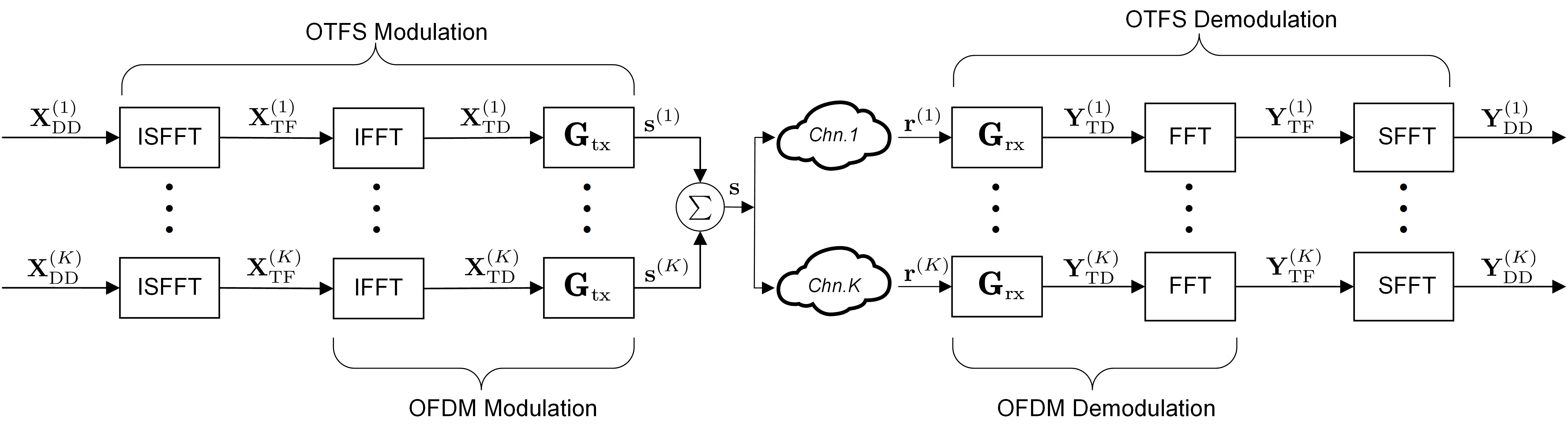}
	\caption{The transmitter structure of MU-SISO-OTFS transmissions.}
	\label{transmitter structure}
\end{figure*}

Consider a MU-SISO LEO satellite communication system serving $K$ users. 
We assume that the resource grid in the DD domain is $T$ seconds wide along the delay domain and $\Delta f=1/T$ Hz wide along the Doppler domain. The delay and Doppler domain are further divided into $M$ and $N$ bins. 
\rev{Since the OTFS modulation is built on the OFDM modulation, for an OTFS frame, its corresponding OFDM frame mapped within the TF domain occupies $N$ OFDM symbols and $M$ subcarriers for each OFDM symbol. In addition, each symbol has the duration of $T$, and each subcarrier has an equal spacing of $\Delta f=1/T$.}

\subsection{Input-Output Relation of MU-SISO-OTFS Transmissions}
\subsubsection{OTFS Modulation at Transmitter}
Consider the MU-SISO-OTFS system in Fig.~\ref{transmitter structure}.
Let $\mathbf{x}_{\text{DD}}^{(i)} \in \mathbb{A}^{MN\times1}$ be the DD domain information symbol vector with length $MN$ of the $i$-th user.
This symbol vector can be transformed into a two-dimensional (2D) symbol matrix $\mathbf{X}_{\text{DD}}^{(i)} \in \mathbb{A}^{M\times N}$ through $\mathbf{x}_{\text{DD}}^{(i)}=\mathrm{vec}\left(\mathbf{X}_{\text{DD}}^{(i)}\right)$. $X_{\text{DD}}^{(i)}\left[l,k\right]$, the $\left(l,k\right)$-th element of $\mathbf{X}_\text{{DD}}^{(i)}$, is the symbol at the $l$-th delay bin and the $k$-th Doppler bin, for $0\le l < M-1,\; 0\le k< N-1$ \cite{2017OTFSFundamentals}. 

Via an inverse symplectic finite Fourier transform (ISFFT) \cite{input-outputRelation}, $\mathbf{X}_{\text{DD}}^{(i)}$ is transformed into the corresponding TF domain symbol matrix $\mathbf{X}_{\text{TF}}^{(i)}$, i.e., 
\begin{equation}
	\mathbf{X}_{\text{TF}}^{(i)}=\mathbf{F}_{M}\mathbf{X}_{\text{DD}}^{(i)}\mathbf{F}_{N}^{\dag},
\end{equation}
where $\mathbf{F}_M$ and $\mathbf{F}_N$ are the normalized $M$-point DFT matrix and $N$-point DFT matrix, respectively.
Similarly, the $(m,n)$-th element of $\mathbf{X}_{\text{TF}}^{(i)}$, $X_{\text{TF}}^{(i)}[m,n]$, is the symbol at the $m$-th subcarrier and $n$-th time slot, for $0 \le m <M-1, \; 0\le n < N-1$ \cite{2017OTFSFundamentals}.
$\mathbf{x}_{\text{TF}}^{(i)}$ is the corresponding vector form of $\mathbf{X}_{\text{TF}}$ given by \cite{2022CrossDomainDetection},
\begin{equation}
	\mathbf{x}_{\text{TF}}^{(i)}=\mathrm{vec}\left(\mathbf{X}_{\text{TF}}^{(i)}\right)\triangleq\left(\mathbf{F}_{N}^{\dagger}\otimes \mathbf{F}_{M}\right)\mathbf{x}_{\text{DD}}^{(i)}.
\end{equation}

Next, the Heisenberg transform is employed on $\mathbf{x}_{\text{TF}}^{(i)}$ to obtain the time-delay (TD) domain symbol vector of the $i$-th user $\mathbf{x}_{\text{TD}}^{(i)}$ \cite{input-outputRelation},
\begin{equation}
	\mathbf{x}^{(i)}_{\text{TD}}=\left(\mathbf{I}_N \otimes \mathbf{F}_M^{\dagger}\right)\mathbf{x}^{(i)}_{\text{TF}}=\left(\mathbf{F}_N^{\dagger}\otimes \mathbf{I}_M\right)\mathbf{x}^{(i)}_{\text{DD}}.
\end{equation}
where $\mathbf{I}_M$ and $\mathbf{I}_N$ are the identity matrices of size $M\times M$ and $N \times N$. $\mathbf{X}_{TD}^{(i)}$ is the corresponding matrix form of $\mathbf{x}_{\text{TD}}^{(i)}$, which follows $\mathbf{x}_{\text{TD}}^{(i)}=\mathrm{vec}\left(\mathbf{X}_{\text{TD}}^{(i)}\right)$.

After the transmitter pulse shaping of the TD domain symbol matrix $\mathbf{X}_{\text{TD}}^{(i)}$, the time domain samples are obtained via 
\begin{equation}
	\mathbf{s}=\sum_{i=1}^{K}\mathbf{s}^{(i)}=\sum_{i=1}^{K}\mathrm{vec}\left(\mathbf{G}_{\text{tx}}\mathbf{X}_{\text{TD}}^{(i)}\right),
\end{equation}
where $\mathbf{G}_{\text{tx}}$ has the the samples of the pulse $g_{\text{tx}}(t)$ as its entries,
\begin{equation}
	\mathbf{G}_{\text{tx}}=\mathrm{diag}\left[g_{\text{tx}}(0),g_{\text{tx}}(T/M),\cdots,g_{\text{tx}}((M-1)T/M)\right].
\end{equation}
For the rectangular waveform, $\mathbf{G}_{\text{tx}}=\mathbf{I}_M$.

\subsubsection{LTV Wireless Channel}
Next, the LTV wireless channel for MU-SISO transmissions should be modeled. Consider the far field assumption \cite{FundamentalsofWirelessCommunication}, the DD domain channel for the $i$-th user can be modeled as
\begin{equation}
	h(i,\tau,\upsilon)=\sum_{p=1}^{P}h_p^{(i)}\delta\left(\tau-\tau_p^{(i)}\right)\delta\left(\upsilon-\upsilon_p^{(i)}\right).
	\label{MU DD domain channel response}
\end{equation}
In \eqref{MU DD domain channel response}, $h_p^{(i)} \in \mathbb{C}$, $\tau_p^{(i)}$ and $\upsilon_p^{(i)}$ are the fading coefficient, delay and Doppler shift for the $p$-th path of the $i$-th user, respectively. 
For the LEO satellite communication channel, the delays and Doppler shifts are modeled according to 3GPP standards for non-terrestrial communications \cite{3GPPTR38.811}.
The original delay $\tau_p^{(i)}$ is given by $\tau_p^{(i)}=\tau_{p,\text{norm}}^{(i)}\cdot\tau_{\text{spread}}$, where $\tau_{\text{spread}}$ is the delay spread and $\tau_{p,\text{norm}}^{(i)}$ the normalized delay. $\upsilon_p^{(i)}$ satisfies $\upsilon_p^{(i)} \le f_d$, where $f_d$ is the maximum Doppler shift, whose values in the distinct bands are given in Table. 5.3.5-1 in \cite{3GPPTR38.811}.

The delay and Doppler shift indices for the $p$-th path of the $i$-th user in \eqref{MU DD domain channel response} are given by 
\begin{equation}
	\tau_p^{(i)}=\frac{l_p^{(i)}}{M\Delta f}, \; \upsilon^{(i)}_p=\frac{k_p^{(i)}}{NT}.
\end{equation}
According to the literature, the effect of fractional delay and Doppler-shift can be mitigated via adding TF domain windows \cite{2021TFdomainwindow} or message passing detection algorithm \cite{2018Iterativedetection}. Henceforth, for ease of derivations, we assume that the delay and Doppler shifts as integer multipliers of $\frac{1}{M\Delta f}$ and $\frac{1}{NT}$ respectively. 

\subsubsection{MU-OTFS Demodulation at Receiver}
It is assumed that the maximum delay of the channel is $\tau_{\rm max}=(M-1)T/M$, which means $\mathrm{max}(l_p)=M-1$. The received signal is sampled at a rate $f_s=M\Delta f=M/T$ and, after discarding the reduced cyclic prefix (RCP), a received samples vector $\mathbf{r}^{(i)}$ is formed, 
\begin{equation}
	\mathbf{r}^{(i)}=\mathbf{H}^{(i)}_{\text{TD}}\mathbf{s} +\mathbf{w}^{(i)}_{\text{TD}},
	\label{TD input-output relation}
\end{equation}
where $\mathbf{w}_{\text{TD}}^{(i)}$ is the corresponding AWGN sample vector in the TD domain with one-sided power spectral density (PSD) $N_0$ and $\mathbf{H}_{\text{TD}}^{(i)}$ is a $MN\times MN$ matrix given by
\begin{equation}
	\mathbf{H}^{(i)}_{\text{TD}}=\sum_{p=1}^{P}h_p^{(i)}\bm{\Pi}^{l_p^{(i)}}\bm{\Delta}^{k_p^{(i)}}.
	\label{H_TD}
\end{equation}
In \eqref{H_TD}, $\bm{\Pi}$ is the permutation matrix (forward cyclic shift) defined as
\begin{equation}
	\bm{\Pi}=\begin{bmatrix}
		0&		\cdots&		0&		1\\
		1&		\ddots&		0&		0\\
		\vdots&		\ddots&		\ddots&		\vdots\\
		0&		\cdots&		1&		0\\
	\end{bmatrix}_{MN\times MN} \quad, 
\end{equation}
and $\bm{\Delta}$ is a $MN\times MN$ diagonal matrix, i.e.,
\begin{equation}
	\bm{\Delta}=\mathrm{diag}\{\gamma^0,\gamma^1,\cdots,\gamma^{MN-1}\},
\end{equation} 
with $\gamma=e^{j\frac{2\pi}{MN}}$. Here, each path introduces an $l_p^{(i)}$-step cyclic shift and $k_p^{(i)}$-step a frequency offset of the transmitted signal vector $\mathbf{s}$, modeled by $\bm{\Pi}^{l_p^{(i)}}$ and $\bm{\Delta}^{k_p^{(i)}}$ respectively.

With the receiver pulse shaping matrix $\mathbf{G}_{\text{rx}}$, the time domain received samples $\mathbf{r}^{(i)}$ is converted to the TD domain matrix $\mathbf{Y}_{\text{TD}}^{(i)}$ via
\begin{equation}
	\mathbf{Y}^{(i)}_{\text{TD}}=\mathbf{G}_{\text{rx}}\left(\mathrm{vec}^{-1}_{M,N}\left(\mathbf{r}^{(i)}\right)\right),
\end{equation}
where $\mathbf{G}_{\text{rx}}$ has the the samples of the pulse $g_{\text{rx}}(t)$ as its entries,
\begin{equation}
	\mathbf{G}_{\text{rx}}=\mathrm{diag}\left[g_{\text{rx}}(0),g_{\text{rx}}(T/M),\cdots,g_{\text{rx}}((M-1)T/M)\right].
\end{equation}
For the rectangular waveform, $\mathbf{G}_{\text{rx}}=\mathbf{I}_M$. $\mathbf{y}_{\text{TD}}^{(i)}$ is the corresponding vector form of $\mathbf{Y}_{\text{TD}}^{(i)}$, following $\mathbf{y}_{\text{TD}}^{(i)}=\mathrm{vec}\left(\mathbf{Y}_{\text{TD}}^{(i)}\right)$.

Then, the received signal vector within the TF domain, $\mathbf{y}^{(i)}_{\text{TF}}$, is obtianed via the Wigner transform,
\begin{equation}
	\mathbf{y}_{\text{TF}}^{(i)} = \mathbf{y}_{\text{TD}}^{(i)}\left(\mathbf{I}_N\otimes\mathbf{F}_M\right),
	\label{y_TF}
\end{equation}
and its matrix form $\mathbf{Y}_{\text{TF}}^{(i)}$ follows $\mathbf{Y}_{\text{TF}}^{(i)}=\mathrm{vec}_{M,N}^{-1}\left(\mathbf{y}_{\text{TF}}^{(i)}\right)$.
Next, through SFFT on $\mathbf{y}_{\text{TF}}^{(i)}$, $\mathbf{y}^{(i)}_{\text{DD}}$, the received signal vector within the DD domain, is given by
\begin{equation}
	\mathbf{y}^{(i)}_{\text{DD}}=\left(\mathbf{F}_N\otimes\mathbf{F}_M^{\dagger}\right)\mathbf{y}^{(i)}_{\text{TF}},
\end{equation}
and $\mathbf{Y}_{\text{DD}}^{(i)}$ is the matricization of $\mathbf{y}_{\text{DD}}^{(i)}$, i.e., $\mathbf{Y_{\text{DD}}}^{(i)}=\mathrm{vec}_{M,N}^{-1}\left(\mathbf{y}_{\text{DD}}^{(i)}\right)$.
\subsubsection{Input-Output Relation}
Henceforth, the OTFS demodulation can be interpreted as the concatenation of the Wigner transform and the SFFT \cite{2018OTFSFundamentals}. Based on \eqref{TD input-output relation}, the received vector of the $i$-th user within the DD domain, $\mathbf{y}_{\text{DD}}^{(i)}$, is given by
\begin{equation}
	\mathbf{y}_{\text{DD}}^{(i)}=\mathbf{H}^{(i)}_{\text{DD}}\mathbf{x}^{(i)}_{\text{DD}}+\mathbf{w}_{\text{DD}}^{(i)},
	\label{DD input-output relation}
\end{equation}
where $\mathbf{H}_{\text{DD}}^{(i)}$ is the corresponding effective DD domain channel matrix for the $i$-th user given by \cite{input-outputRelation}
\begin{equation}
	\mathbf{H}^{(i)}_{\text{DD}}=\sum_{p=1}^{P}h_p^{(i)}\left(\mathbf{F}_N\otimes\mathbf{I}_M\right)\bm{\Pi}^{l_p^{(i)}}\bm{\Delta}^{k_p^{(i)}}\left(\mathbf{F}_N^{\dagger}\otimes\mathbf{I}_M\right).
	\label{H_DD}
\end{equation}

\subsection{Derivation of Achievable Rate for OTFS}
In this subsection, we derive the achievable rate.
First, the DD domain interference pattern is analyzed to provide some insights. For ease of derivation, the DD domain input-output relation \eqref{DD input-output relation} can be rewritten into a symbol-wise form via the inverse discrete Zak transform (IDZT).
According to the derivation in Chapter 5 of \cite{hong2022delayDoppler}, with the rectangular pulse shaping waveform and RCP structure, the $(l,k)$-th entry of $\mathbf{Y}_{\text{DD}}^{(i)}$ is given by
\begin{equation}
	\begin{aligned}
		Y_{\text{DD}}^{(i)}[l,k]=&\sum_{p=1}^{P}\sum_{j=1}^{K}g_{l,k}^{i,p}X_{\text{DD}}^{(j)}\left[[l-l_p^{(i)}]_M,[k-k_p^{(i)}]_N\right] \\
		& + W_{\text{DD}}^{(i)}[l,k],
	\end{aligned}
	\label{MU symbol-wise input out relation}
\end{equation} 
where $g_{l,k}^{i,p}$ characterizes the symbol-wise effective channel coefficient given by
\begin{equation}
	g_{l,k}^{i,p}=\begin{cases}
		h_p^{(i)}e^{j2\pi\frac{k_p^{(i)}\left(l-l_p^{(i)}\right)}{MN}}, & \; l-l_p^{(i)}\ge 0 \\
		h_p^{(i)}e^{j2\pi\frac{k_p^{(i)}\left[l-l_p^{(i)}\right]_M}{MN}}e^{-j2\pi\frac{k}{N}}, & \; l-l_p^{(i)} < 0
	\end{cases}
\end{equation}
and $W_{\text{DD}}^{(i)}$ is the DD domain noise with one side PSD $N_0$.
Assuming that the first-tap path carries the desired signal for the transmission, we can expand \eqref{MU symbol-wise input out relation} to yield
	\begin{equation}
		\begin{aligned}
			Y_{\text{DD}}^{(i)}[l,k] = &
			\underbrace{g_{l,k}^{i,1}X_{\text{DD}}^{(i)}\left[[l-l_1^{(i)}]_M,[k-k_1^{(i)}]_N\right]}_{\text{Desired signal}} \\
			& + \underbrace{\sum_{p=2}^{P}g_{l,k}^{i,p}X_{\text{DD}}^{(i)}\left[[l-l_p^{(i)}]_M,[k-k_p^{(i)}]_N\right]}_{\text{MPSI}}\\
			& + \underbrace{\sum_{j\neq i}\sum_{p=1}^{P}g_{l,k}^{i,p}X_{\text{DD}}^{(j)}\left[[l-l_p^{(i)}]_M,[k-k_p^{(i)}]_N\right]}_{\text{MUI}} \\
			& + \underbrace{W_{\text{DD}}^{(i)}[l,k]}_{\text{noise}}.
		\end{aligned}
		\label{MU symbol-wise input out relation Expanded}
	\end{equation}
From \eqref{MU symbol-wise input out relation Expanded}, it is noted that $Y_{\text{DD}}^{(i)}$ is composed of the desired signals containing information, noise, and the diverse interference defined as follows,
\begin{itemize}
	\item The second term in \eqref{MU symbol-wise input out relation Expanded} is the MPSI, which is caused by the multipath transmissions via the non-main path of the desired user. 
	\item The third term in \eqref{MU symbol-wise input out relation Expanded} is the MUI, which is caused by the transmissions of other users.
\end{itemize}

Next, we derive the expression of the achievable rate of the transmission system. 
For ease of derivation, we consider that the symbols are i.i.d., and follow a Gaussian distribution $X_{\text{DD}}^{(i)}[l,k]\sim\mathcal{CN}(0,\rho_{l,k}^{(i)})$. Due to the fact that $X^{(i)}_{\text{DD}}[l,k]$ and $W_{\text{DD}}^{(i)}[l,k]$ are independent with each other, the signal-to-interference-plus-noise ratio (SINR) of $Y_{\text{DD}}^{(i)}[l,k]$, the $(l,k)$-th received symbol for the $i$-th user, can be given by \eqref{SINR expression}, where $\left|h_p^{(i)}\right|^2$=$\left|g_{l,k}^{i,p}\right|^2$.

According to \cite{2006Elements}, the the achievable rate of the $(l,k)$-th symbol for the $i$-th user is given by
\begin{equation}
	R_{l,k}^{(i)}=\frac{1}{2}\log_2\left(1+\Gamma_{l,k}^{(i)}\right).
\end{equation}
Consequently, the sum rate of the system is expressed as
\begin{equation}
	R=\sum_{i=1}^{K}\sum_{l=0}^{M-1}\sum_{k=0}^{N-1}R_{l,k}^{(i)}.
\end{equation}

\begin{figure*}[tb]
\vspace*{-20pt}
\begin{equation}
	\Gamma_{l,k}^{(i)}=\frac{\left|h_1^{(i)}\right|^2\rho_{[l-l_1^{(i)}]_M,[k-k_1^{(i)}]_N}^{(i)}}
	{\sum\limits_{p=2}^{P}\left|h_p^{(i)}\right|^2\rho_{[l-l_p^{(i)}]_M,[k-k_p^{(i)}]_N}^{(i)}+\sum\limits_{j\neq i}\sum\limits_{p=1}^P\left|h_p^{(i)}\right|^2\rho_{[l-l_p^{(i)}]_M,[k-k_p^{(i)}]_N}^{(j)}+N_0}
	\label{SINR expression}
\end{equation}
\vspace*{-15pt}
\rule{\linewidth}{.5pt}
\end{figure*}

\section{Problem Formulation}
In a practical multiuser OTFS system, symbols in the transmitted symbol matrix suffer from strong interference from the neighboring symbols due to the impacts of multipaths and Doppler shfts, which results in poor received SINR and quality of service. Simply increasing the transmission power for one symbol does not necessarily improve the overall system performance as interference to other symbols increases as well. Therefore, a practical multi-symbol scheduler should be able to coordinate the strength of interference caused by each symbol. Some existing multi-access methods are listed in Appendix A. For an optimal joint power allocation and multiuser's symbol scheduling policy, we organize the following optimization problem. 

Define an $M\times N\times K$ 3-dimensional (3D) matrix $\bm{\rho}$ as the power allocation policy, in which $\rho_{l,k}^{(i)}$ determines the power allocated to the symbol in the $(l,k)$-th resource block for the $i$-th user. Define an $M\times N\times K$ 3D matrix $\pmb{S}$ as the multiuser symbol scheduling policy, in which $s_{l,k}^{(i)}\in \{0,1\}$ determines whether the symbol in the $(l,k)$-th resource block is scheduled for the $i$-th user, which follows $s_{l,k}^{(i)}=\mathrm{sgn}\left(\rho_{l,k}^{(i)}\right)$. 
It is evident that the sum rate $R$ is intrinsically dependent on the power allocation and symbol scheduling. 
To maximize the sum rate $R$, the optimal power allocation policy $\bm{\rho}^*$ and symbol scheduling policy $\pmb{S}^*$ are the solution of the optimization problem (P1),  
\begin{equation}
\notag
\begin{aligned}
	\textbf{(P1)}\qquad \max_{\bm{\rho},\pmb{S}} \quad & R \\
	\text{s.t.} \quad & \text{C1:\;}  \sum_{i=1}^K\sum_{l=0}^{M-1}\sum_{k=0}^{N-1}\rho_{l,k}^{(i)}\le P_0 \\
	 & \text{C2:\;}  \rho_{l,k}^{(i)} \ge 0, \; \forall l,k,i \\
	 & \text{C3:\;}  s_{l,k}^{(i)}=\mathrm{sgn}\left(\rho_{l,k}^{(i)}\right), \; \forall l,k,i \\
	 & \text{C4:\;}  \sum_{i=1}^{K} s_{l,k}^{(i)} \le 1, \; \forall l,k.
\end{aligned}
\label{original problem}
\end{equation}
Here, C1 represents the total power constraint for the LEO satellite with maximum transmit power $P_0$. C2 is the positive power constraint, C3 enforces $s_{l,k}^{(i)}$ associated with the allocated power $\rho_{l,k}^{(i)}$. C4 guarantees that each resource block is only occupied by one symbol. 

\section{DD Domain Resource Allocation Algorithm}
In this section, we propose a DD domain power allocation and symbols scheduling algorithm to solve the problem (P1).
In particular, we assume that the channel state information (CSI) is available at the transmitter, which can be achieved by exploiting the DD domain reciprocity based on uplink channel estimation \cite{2021OTFSPDMA}.
The considered problem is a mixed combinatorial and non-convex optimization problem. The combinatorial nature comes from the conditional constraint for the resource blocks occupation strategy for symbols while the non-convexity is caused by MPSI and MUI in \eqref{MU symbol-wise input out relation Expanded}. In general, a brute force approach is needed to obtain the globally optimal solution. In an OTFS multiuser system where single resource grid composed of $M$ delay domain bins, $N$ Doppler-shift bins and $K$ users, there are $M^{N^K}$ possible symbol assignments which limit the scalability in practical systems. In order to make the problem tractable, we perform the following transformation to simplify the problem.
\begin{enumerate}
	\item We relax the discrete indicator variables $s_{l,k}^{(i)}$ to the continuous optimization variables via transforming the conditional constraint C3 in (P1) into a series of equivalent constraints in (P2).
	\item We employ the penalty convex concave procedure (CCP) algorithm on the difference of convex (DC) problem (P2), which transform the solution of (P2) into an equivalent process of solving a sequence of convex subproblems (P3).
\end{enumerate}

\subsection{Transformation of Optimization Problem}
The first step in solving the considered problem is to handle the conditional constraint C3, which restricts the variables $s_{l,k}^{(i)}$ to binary integer values of ${0,1}$.
To this end, we introduce the Big-M method to transform the conditional constraint C3 into the following equivalent constraints,
\begin{subequations}
\begin{align}
	\rho_{l,k}^{(i)} \le& P_0 s^{(i)}_{l,k} , \\
	\rho_{l,k}^{(i)} \ge& P_0 \left(s_{l,k}^{(i)}-1\right) + \epsilon , \\
	s_{l,k}^{(i)} \in& \{0,1\}. \label{combinatorial constraint}
\end{align}
\end{subequations}
where $\epsilon$ has a sufficiently small positive value. 
Furthermore, in order to deal with the combinatorial constraint \eqref{combinatorial constraint}, we transform it into 
\begin{equation}
	s_{l,k}^{(i)}\left(s_{l,k}^{(i)}-1\right)=0.
\end{equation}
Henceforth, the resource and scheduling optimization problem (P1) can be transformed into 
\begin{equation*}
\begin{aligned}
	\textbf{(P2)} \qquad \max_{\bm{\rho},\pmb{S}} \quad & R \\
	\text{s.t.} \quad & \text{C1, C2, C4} \\
	&  \text{C3:\;} \rho_{l,k}^{(i)} \le P_0 s^{(i)}_{l,k}, \; \forall l,k,i \\
	&  \text{C5:\;} \rho_{l,k}^{(i)} \ge P_0 \left(s^{(i)}_{l,k}-1\right)+\epsilon, \; \forall l,k,i \\
	&  \text{C6:\;} s_{l,k}^{(i)}\left(s_{l,k}^{(i)}-1\right)\le 0, \; \forall l,k,i \\
	&  \text{C7:\;} s_{l,k}^{(i)}\left(s_{l,k}^{(i)}-1\right)\ge 0, \; \forall l,k,i .
\end{aligned}
\end{equation*}
Now the objective function follows a convex-concave form and the constraints are convex except C6. 
Next, we employ the penalty CCP algorithm to solve the considered problem.

\subsection{Penalty CCP Algorithm}
In this subsection, the transformed power allocation and symbol scheduling optimization problem is solved by the penalty CCP algorithm. For this purpose, we firstly need to rewrite the objective function into a DC form: 
\begin{equation}
	R=\sum_{i=1}^K\sum_{l=0}^{M-1}\sum_{k=0}^{N-1}\left(Q_{l,k}^{(i)}-Z_{l,k}^{(i)}\right)
	\triangleq \bar{Q}-\bar{Z}
\end{equation}
where 
\begin{equation}
	\!\!\!
	Q_{l,k}^{(i)}=\log_2\left(\sum_{j=1}^K\sum_{p=1}^P\Bigl|h_p^{(i)}\Bigr|^2\rho_{[l-l_p^{(i)}]_M,[k-k_p^{(i)}]_N}^{(j)}+N_0\right),
\end{equation}
\begin{equation}
\begin{aligned}
	Z_{l,k}^{(i)}=&\log_2\left(\sum_{p=2}^P\Bigl|h_p^{(i)}\Bigr|^2\rho_{[l-l_p^{(i)}]_M,[k-k_p^{(i)}]_N}^{(i)} \right. \\
	&\; \left. +\sum_{j\neq i}\sum_{p=1}^P\Bigl|h_p^{(i)}\Bigr|^2\rho_{[l-l_p^{(i)}]_M,[k-k_p^{(i)}]_N}^{(j)}+N_0\right),
\end{aligned}
\end{equation}
$\bar{Q}$ and $\bar{Z}$ are the sum of $Q_{l,k}^{(i)}$ and $Z_{l,k}^{(i)}$, respectively.

Due to the modulo operation involved, we should establish a relation between $(l,k)$, the index of the resource block at the transmitter, and $(\tilde{l},\tilde{k})$, the index of the resource block at the receiver. With the offsets $l_p^{(i)}$ and $k_p^{(i)}$, we have the following relation as
\begin{equation}
	l=[\tilde{l}+l_p^{(i)}]_M, \quad k=[\tilde{k}+k_p^{(i)}]_N.
\end{equation}

To employ the penalty CCP algorithm, we should apply first-order Taylor expansion on the concave term $Z_{l,k}^{(i)}$. 
Assume that $\rho_{l,k}^{(i),(m)}$ and $\bm{\rho}^{(m)}$ are the values of $\rho_{l,k}^{(i)}$ and $\bm{\rho}$, respectively, in the $m$-th iteration of the algorithm. Considering that the logarithmic function is equal to or smaller than its first Taylor expansion at any point, we have
\begin{equation}
\begin{aligned}
	\bar{Z} & \le \bar{Z}^{(m)}+\sum_{i=1}^{K}\sum_{l=0}^{M-1}\sum_{k=0}^{N-1}\frac{\partial \bar{Z}}{\partial \rho_{l,k}^{(i)}}\cdot \left(\rho_{l,k}^{(i)}-\rho_{l,k}^{(i),(m)}\right) \\
	& \triangleq \hat{Z}^{(m)},
\end{aligned}
\end{equation} 
where $\bar{Z}^{(m)}$ is $\bar{Z}$ with respect to $\bm{\rho}=\bm{\rho}^{(m)}$, and the partial derivative $\partial \bar{Z}/\partial \rho_{l,k}^{(i)}$ is given by \eqref{derivative expression}.
\begin{figure*}[tb]
\vspace*{-20pt}
\begin{equation}
		\frac{\partial\bar{Z}}{\partial\rho_{l,k}^{(i)}}=\frac{1}{\ln2}\sum_{p'=2}^{P}\frac{\bigl|h_{p'}^{(i)}\bigr|^2}{\sum\limits_{p=2}^P\bigl|h_{p'}^{(i)}\bigr|^2\rho_{\bigl[[l+l_{p'}^{(i)}]_M-l_p^{(i)}\bigr]_M,\bigl[[k+k_{p'}^{(i)}]_N-k_p^{(i)}\bigr]_N}^{(i),(m)}+\sum\limits_{j\neq i}\sum\limits_{p=1}^P\bigl|h_{p'}^{(i)}\bigr|^2\rho_{\bigl[[l+l_{p'}^{(i)}]_M-l_p^{(i)}\bigr]_M,\bigl[[k+k_{p'}^{(i)}]_N-k_p^{(i)}\bigr]_N}^{(j),(m)}+N_0}
		\label{derivative expression}
\end{equation}
\vspace*{-15pt}
\rule{\linewidth}{.5pt}
\end{figure*}

To ensure \rev{that} the constraints of (P2) convex, the concave constraint C7 should be expanded via the first-order Taylor expansion as 
\begin{multline}
	\left(s_{l,k}^{(i),(m)}-\left(s_{l,k}^{(i),(m)}\right)^2\right) \\ 
	+\left(1-2s_{l,k}^{(i),(m)}\right)\left(s_{l,k}^{(i)}-s_{l,k}^{(i),(m)}\right) \le a_{l,k}^{(i)}, \;\forall l,k,i
	\label{s;first-order Taylor Expansion}
\end{multline}
where $s_{l,k}^{(i),(m)}$ is the $m$-th iteration of $s_{l,k}^{(i)}$ in the penalty CCP algorithm and $a_{l,k}^{(i)}$ is the slack variables used to relax the problem. Henceforth, relaxed through adding slack variables, the solution of (P2) can be transformed into solving the following subproblems (P3) sequentially,
\begin{equation*}
\begin{aligned}
	\textbf{(P3)} \quad \max_{\bm{\rho},\pmb{S},\pmb{A}} & \; \bar{Q}\left(\bm{\rho}\right)-\hat{Z}\left(\bm{\rho};\bm{\rho}^{(m)}\right)-\xi^{(m)}\sum_{i=1}^K\sum_{l=0}^{M-1}\sum_{k=0}^{N-1}a_{l,k}^{(i)} \\
	\text{s.t.} &  \; \text{C1, C2, C3, C4, C5, C6} \\
	& \; \text{C7: } \eqref{s;first-order Taylor Expansion} \\
	& \; \text{C8: } a_{l,k}^{(i)} \ge 0, \; \forall l,k,i
\end{aligned}
\end{equation*}
where $\xi^{(m)}$ is the regularization factor to scale the impact of the penalty term $\sum_{i=1}^K\sum_{l=0}^{M-1}\sum_{k=0}^{N-1}a_{l,k}^{(i)}$, which controls the feasibility of the constraints. 
When the values of $\xi^{(m)}$ are small, (P3) reaches the maximum rate while its solution can be infeasible for (P1). When the values of $\xi^{(m)}$ become large, (P3) seeks for a feasible point for the original problem. 

(P3) is convex and can be solved by CVX tool. The algorithm for finding a feasible solution of $\bm{\rho}$ and $\pmb{A}$ is summarized in Algorithm \ref{penalty CCP}. Some points are emphasized as follows: 
\begin{enumerate}[a)]
  \item The maximum penalty scaling parameter $\xi_{\rm max}$ is imposed to avoid the numerical results, i.e., a feasible solution may not be found when the iteration converges under increasing large values of $\xi^{(m)}$ 
  \item The stopping criterion $\left\|\pmb{A}^{(m+1)}-\pmb{A}^{(m)}\right\|_1\le\delta_2$ guarantees the binary variable constraint C7 in (P3) to be met for a sufficiently low $\delta_2$. 
  \item The stopping criterion $\left\|\bm{\rho}^{(m+1)}-\bm{\rho}^{(m)}\right\|_1\le \delta_1$ controls the convergence of the CCP algorithm.
  \item As mentioned in \cite{lipp2016PenaltyCCP}, a feasible solution for (P3) may not be feasible for \textit{Problem 2}. Henceforth, according to \cite{2014DCA}, the feasibility of \textit{Problem 2} can be guaranteed by imposing a maximum number of iterations $m_{\rm max}$. \cite{2014DCA} proves that when $m$ is finite and sufficiently large, $\|\mathbf{A}^{(m)}\|_1=0$. Henceforth, if $m_{\rm max}$ is reached, the iteration will be stopped.
               
\end{enumerate}
\begin{algorithm}[htbp]
	\caption{Penalty CCP Algorithm for Power and Access Allocation}
	\label{penalty CCP}
	\KwIn{Take an initial point $\bm{\rho}^{(0)}$; an initial penalty scaling parameter $\xi^{(0)}>0$; a maximum penalty scaling paramter $\xi_{\rm max}$; an update parameter $\mu \ge 1$; stopping criterion $\{\delta_1,\delta_2\}$; and set iteration counter $m\coloneqq 0$}
	\BlankLine
	\Repeat{stopping criterion is satisfied: $\left(\left\|\bm{\rho}^{(m+1)}-\bm{\rho}^{(m)}\right\|_1\le \delta_1 \; \mathrm{and} \; \left\|\pmb{A}^{(m+1)}-\pmb{A}^{(m)}\right\|_1\le\delta_2\right)$ \rm{or} $m\ge m_{\rm max}$}
	{
		\textit{Convexity.} Form $\hat{Z}^{(m)}\triangleq \bar{Z}^{(m)}+\sum_{i=1}^{K}\sum_{l=0}^{M-1}\sum_{k=0}^{N-1}\frac{\partial \bar{Z}}{\partial \rho_{l,k}^{(i)}}\cdot \left(\rho_{l,k}^{(i)}-\rho_{l,k}^{(i),(m)}\right)$ for $\forall l,k,i.$\\
		\textit{Solve.} Set the value of $\bm{\rho}^{(m+1)}$ to a solution of (P3). \\
		\textit{Update $\xi$.} $\xi^{(m+1)}\coloneqq \min\left(\mu\xi^{(m)},\xi_{\rm max}\right)$. \\
		\textit{Update iteration.} $m\coloneqq m+1$.
	}
\end{algorithm}

\section{Simulation Results}
In this section, we present representative 
In order to exhibit the performance evaluation and comparison comprehensively, it is imperative to clarify certain system settings first.
\subsection{System Settings}
First, define the symbol-to-noise-ratio as $\mathrm{SNR}=P_0/MNN_0$. 
Moreover, we introduce the normalized CFO, $\varepsilon=f_\text{offset}/\Delta f$, to measure the effect of Doppler shift, which is defined as a ratio of the CFO to the subcarrier spacing \cite{cho2010mimo}.
The normalized CFO is constitute of the integer carrier frequency offset (IFO) and the fractional carrier frequency offset (FFO).
The IFO will cause a cyclic shift $\lfloor \varepsilon \rfloor$ in the receiver, incurring a significant degradation in \rev{performance}. However, the orthogonality among the subcarrier frequency components is not destroyed, and thus, ICI does not occur. On the contrary, the orthogonality among subcarrier frequency components is not maintained any longer due to the FFO, which implies severe amplitude and phase distortion caused by ICI. 
In our simulations, we \rev{mainly} consider the interference from other subcarriers \rev{rather than} the cyclic shift of symbols. Henceforth, we limit $\varepsilon\le 1$ in the simulations.

For obtaining the simulation results, we set the parameters of the system as in Table. \ref{settings}. To fit the reality, we utilize the 3GPP non-terrestrial networks time delay line (NTN-TDL) channel model \cite{3GPPTR38.811} to construct channel profiles in the simulation. Without loss of generality, we employ the power delay profile of both the NTN-TDL-B model and NTN-TDL-D model to simulate the line-of-sight (LOS) channel condition and the non-line-of-sight (NLOS) channel condition, respectively. 
The detailed power delay profiles are listed in Table.~\ref{pdp NTN_TDL}. The delay and Doppler shift indices are assumed to be integer values unless otherwise specified.  

\begin{table}[htbp]
	\centering
	\caption{Parameters of the LEO Satcom System}
	\begin{tabular}{|c|c|}
		\hline
		Parameter & Value \\ \hline
		Number of delay bins $M$ & 64 \\ \hline
		Number of Doppler bins $N$ & 16 \\ \hline
		Number of users $K$ & 4 \\ \hline
		Earth radius $R$ & 6371 km \\ \hline
		Satellite height $R_h$ & 1500 km \\ \hline
		Elevation angle $\alpha$ & 50$^{\circ}$ \\ \hline
		Satellite speed $v_{sat}$ & 7.11 km/s \\ \hline
		Terminal speed $v_t$ & 500km/h \\ \hline
		Carrier frequency $f_C$ & 2 GHz (S-band) \\ \hline
		Subcarrier frequency $\Delta f$ & 15kHz \\ \hline
		Normalized CFO $\varepsilon$ & [0.25-0.5] \\ \hline
	\end{tabular}
	\label{settings}
\end{table}
\begin{table}[htbp]
	\centering
	\caption{The power delay profile of NTN-TDL-B and NTN-TDL-D model}
	\begin{tabular}{|cccc|}
		\hline
		\multicolumn{4}{|c|}{NTN-TDL-B}                                                                                                         \\ \hline
		\multicolumn{1}{|c|}{Tap \#}             & \multicolumn{1}{c|}{Normalized delay} & \multicolumn{1}{c|}{Power(dB)} & Fading distribution \\ \hline
		\multicolumn{1}{|c|}{1}                  & \multicolumn{1}{c|}{0}                & \multicolumn{1}{c|}{0}         & NLOS,Rayleigh       \\ \hline
		\multicolumn{1}{|c|}{2}                  & \multicolumn{1}{c|}{0.7429}           & \multicolumn{1}{c|}{-1.973}    & NLOS,Rayleigh       \\ \hline
		\multicolumn{1}{|c|}{2}                  & \multicolumn{1}{c|}{0.7410}           & \multicolumn{1}{c|}{-4.332}    & NLOS,Rayleigh       \\ \hline
		\multicolumn{1}{|c|}{3}                  & \multicolumn{1}{c|}{5.792}            & \multicolumn{1}{c|}{-11.914}   & NLOS,Rayleigh       \\ \hline
		\multicolumn{4}{|c|}{NTN-TDL-D}                                                                                                         \\ \hline
		\multicolumn{1}{|c|}{Tap \#}             & \multicolumn{1}{c|}{Normalized delay} & \multicolumn{1}{c|}{Power(dB)} & Fading distribution \\ \hline
		\multicolumn{1}{|c|}{\multirow{2}{*}{1}} & \multicolumn{1}{c|}{0}                & \multicolumn{1}{c|}{-0.284}    & LOS, Ricean         \\ \cline{2-4} 
		\multicolumn{1}{|c|}{}                   & \multicolumn{1}{c|}{0}                & \multicolumn{1}{c|}{-11.991}   & NLOS,Rayleigh       \\ \hline
		\multicolumn{1}{|c|}{2}                  & \multicolumn{1}{c|}{0.5596}           & \multicolumn{1}{c|}{-9.887}    & NLOS,Rayleigh       \\ \hline
		\multicolumn{1}{|c|}{3}                  & \multicolumn{1}{c|}{7.3340}           & \multicolumn{1}{c|}{-16.771}   & NLOS,Rayleigh       \\ \hline
	\end{tabular}
	\label{pdp NTN_TDL}
\end{table}
\subsection{Sum-rate with Various Multiple Access schemes}
\begin{figure}[htbp] 
	\centering
	\subfloat[sum-rate performance under the NTN-TDL-B channel model]
	{	
		\includegraphics[width=0.9\linewidth]{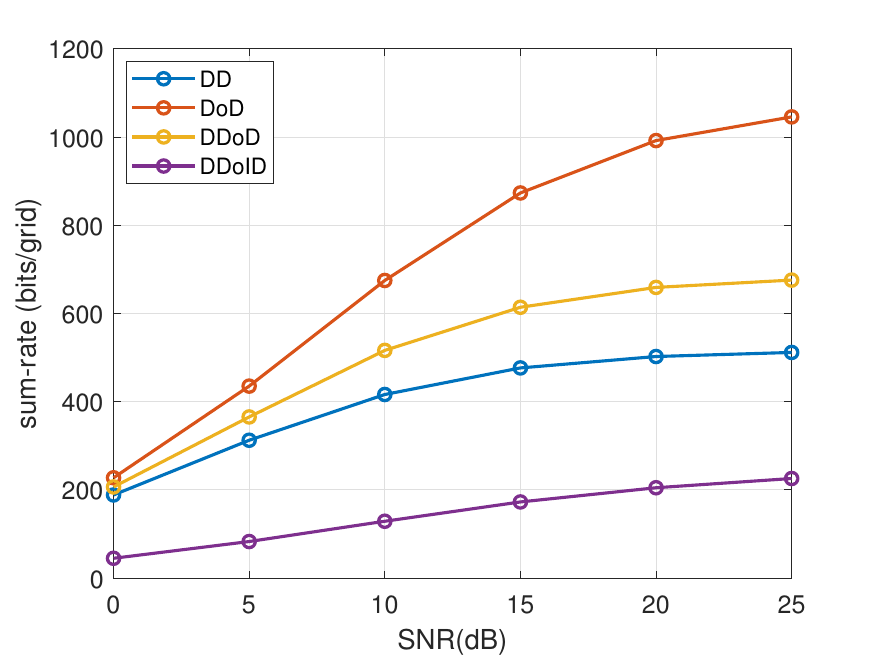}
		\label{sum-rate of OMA NTNT-TDL_B}
		
	}
	\quad
	\subfloat[sum-rate performance under the NTN-TDL-D channel model]
	{
		\includegraphics[width=0.9\linewidth]{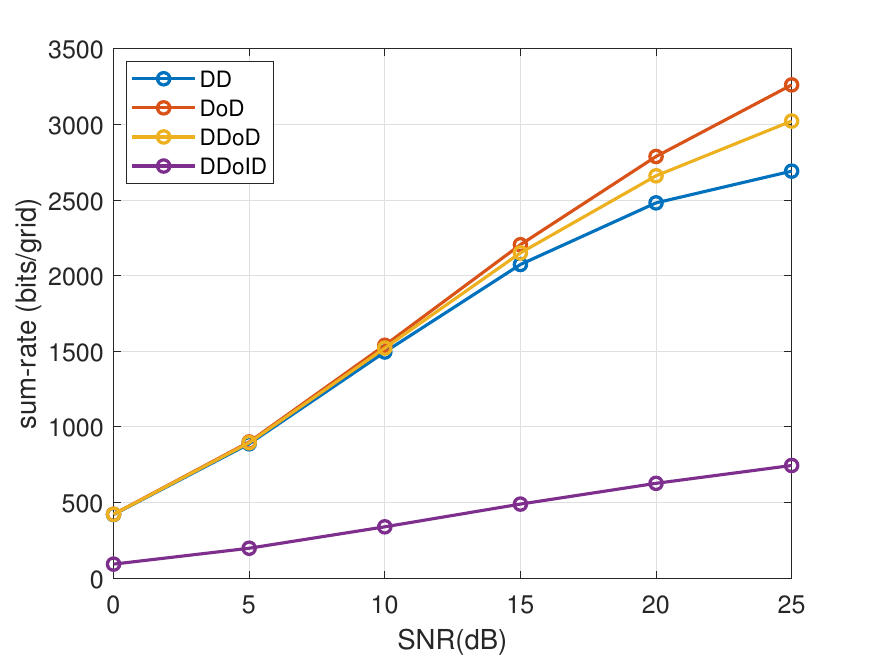}
		\label{sum-rate of OMA NTNT-TDL_D}
	}
	\caption{The sum-rate performances of different orthogonal multiple access schemes with respect to different transmission SNR.}
	\label{sum-rate of OMA}
\end{figure}

In this subsection, we focus on the impact of different multiple access schemes on the system performance. As can be seen from Section III-A and Section III-B, the multiple access schemes, which represent the symbols scheduling, play an important role in the OTFS transmission. The details of four ordinary orthogonal multiple access methods are listed in Appendix A. 
We first present the sum-rate for the MU-OTFS system of different orthogonal multiple access schemes with respect to different transmission SNR in Fig. \ref{sum-rate of OMA}, where we set $\varepsilon=0.25$.
As illustrated in Fig. \ref{sum-rate of OMA}\subref{sum-rate of OMA NTNT-TDL_D}, there is a linear increase in the sum-rate curves with the escalation of transmission SNR under the NTN-TDL-D channel. Concurrently, the sum-rate curves demonstrate a linear ascent in the low SNR regime \rev{with} a marginal plateau in high SNR regime as depicted in Fig. \ref{sum-rate of OMA}\subref{sum-rate of OMA NTNT-TDL_B}.
\rev{Compared} to Fig. \ref{sum-rate of OMA}\subref{sum-rate of OMA NTNT-TDL_D}, it can be seen that the saturation phenomenon in high SNR regime is more obvious in Fig. \ref{sum-rate of OMA}\subref{sum-rate of OMA NTNT-TDL_B}. This can be attributed to the significantly \rev{enhanced} MUI and MPSI, which undermines the potential performance improvement brought by \rev{the increased SNR}. Moreover, it is noted that, in Fig. \ref{sum-rate of OMA}\subref{sum-rate of OMA NTNT-TDL_D}, DDMA, DoDMA and DDoDMA show a superior performance to the DDoIDMA scheme proposed in \cite{2019OTFSMAMohammed}. This is due to \rev{the fact that} the interleaved symbols scheduling may introduce additional interference from adjacent symbols, which degrades the sum rate significantly. Nevertheless, in Fig. \ref{sum-rate of OMA}\subref{sum-rate of OMA NTNT-TDL_B}, it is observed that there exists a marked performance disparity among four ordinary orthogonal multiple access schemes. This observation underscores the profound influence of the channel conditions on the efficacy of multiple access schemes, particularly due to the distinct interference patterns in the DD domain.  
          
\subsection{Sum-rate with Various CFO Settings}
\begin{figure}[htbp]
	\centering
	\subfloat[sum-rate performance with various CFO settings under the NTN-TDL-B channel model]
	{
		\includegraphics[width=0.9\linewidth]{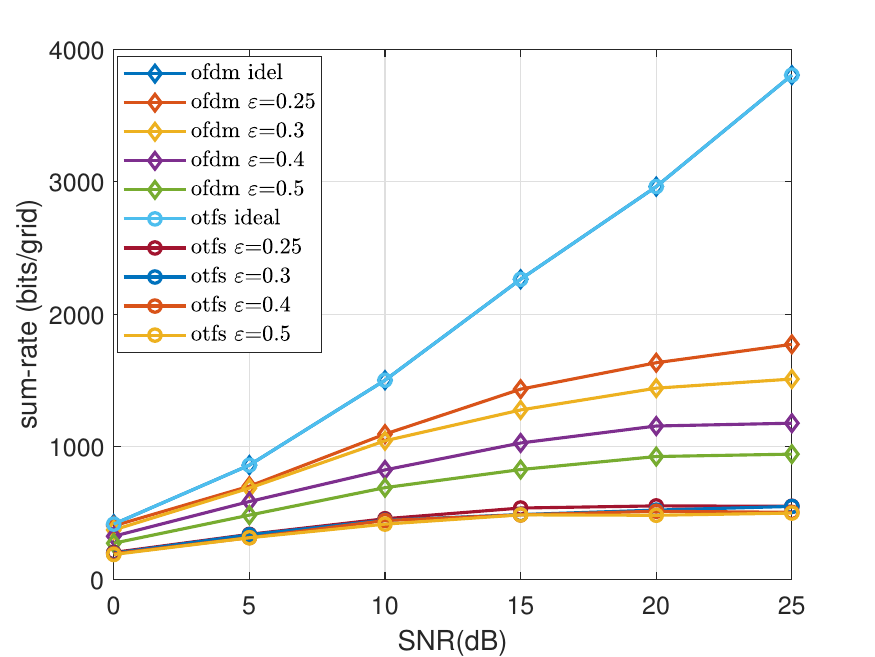}
		\label{sum-rate under CFO NTNT-TDL_B}
	}
	\quad
	\subfloat[sum-rate performance with various CFO settings under the NTN-TDL-D channel model]
	{
		\includegraphics[width=0.9\linewidth]{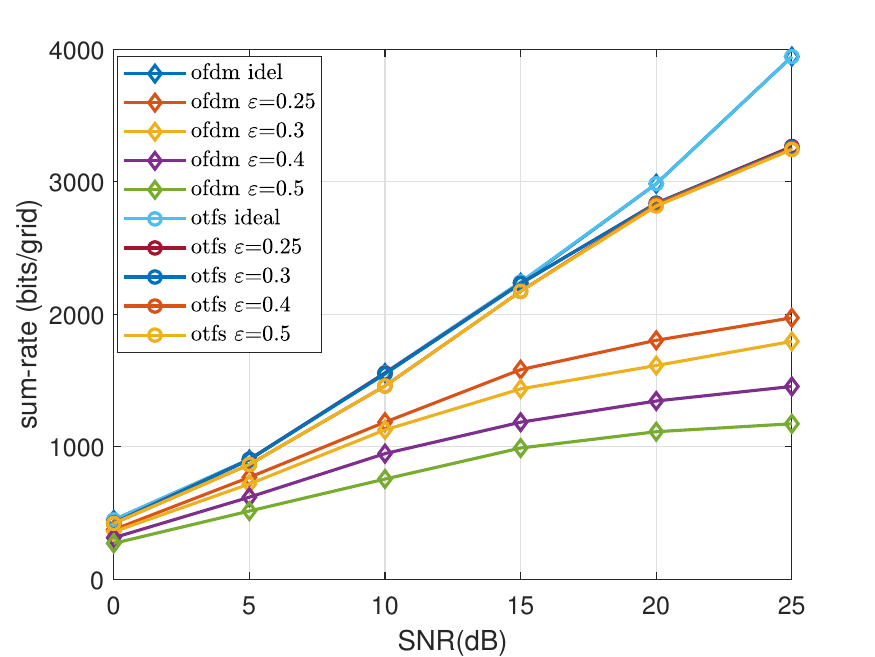}
		\label{sum-rate under CFO NTNT-TDL_D}
	}
	\caption{The sum-rate performances under various CFO settings with respect to different transmission SNR.}
	\label{sum-rate under CFO}
\end{figure}
In this subsection, we focus on the impact of the value of $\varepsilon$ on the system performance. As can be seen from Chapter 5.2 in \cite{cho2010mimo}, the normalized CFO $\varepsilon$, which delineates the extent of Doppler shift caused by high mobility, plays an important role in measuring the ICI. Fig. \ref{sum-rate under CFO} presents a comparative analysis of the sum-rate achieved by the OTFS and OFDM modulation versus the value of $\varepsilon$, under different SNR and channel conditions. 
It is observed that in an ideal channel condition, where UDs are free from multipaths and Doppler shifts, OTFS and OFDM exhibit identical sum-rate performance. 
This similarity indicates that, in the absence of channel influences, both modulations converge towards a shared performance upper bound.  
However, the introduction of multipath effects and Doppler shifts into the analysis highlights a significant divergence in their capabilities.
Specifically, the OFDM modulation is susceptible to the fading effects induced by Doppler shifts. This is evident as the sum-rate of OFDM gradually decreases with increasing values of $\varepsilon$.
On the contrary, the sum-rate of the OTFS modulation maintains robust across different $\varepsilon$ values. This observation demonstrates that, compared to the OFDM modulation, the OTFS modulation exhibits strong robustness over the time-varying channel. 
This phenomenon can be attributed to the capability of delay-Doppler signal representations in capturing the dynamic nature of time-varying channels. This technique, by converting time-delay signal representations to delay-Doppler signal representations via Zak transform, can effectively describe the rapidly and unpredictably alternations of the time-varying channel as several constants in the DD domain, especially in fast fading conditions. Thus, the OTFS modulation is \rev{shown} to ensure the reliability and efficiency of data transmission in high-mobility scenarios.

\subsection{Realistic Transmission Performance with Various CFO Settings}
In this subsection, we further present the realistic transmission performance of OFDM and OTFS modulation, and reveal the relationship between the realistic performance and the achievable sum-rate. 
From Fig. \ref{sum-rate under CFO}, it can be seen that, in NTN-TDL-B channel channel conditions, the OFDM modulation has superior performance to the OTFS modulation on the achievable sum-rate. 
This phenomenon does not imply that the OFDM modulation is superior to the OTFS modulation in practice. 
In order to evaluate the realistic transmission performance of the OFDM and OTFS modulation, we compare the BER under different CFO settings. For the OTFS modulation, we employ the LMMSE method to detect symbols. Moreover, in terms of the OFDM modulation, we employ the one tap method (1tap) with the ideal channel estimation and the ordinary least-square based symbol detection method with the CFO compensation to simulate the ideal condition and the practical condition, respectively. In the simulation, we employ the Moose method to estimate the CFO.
\begin{figure}[htbp]
	\centering
	\subfloat[BER performance of practical OTFS symbol detection and ideal OFDM symbol detection with various CFO settings under the NTN-TDL-B channel model]
	{
		\includegraphics[width=0.9\linewidth]{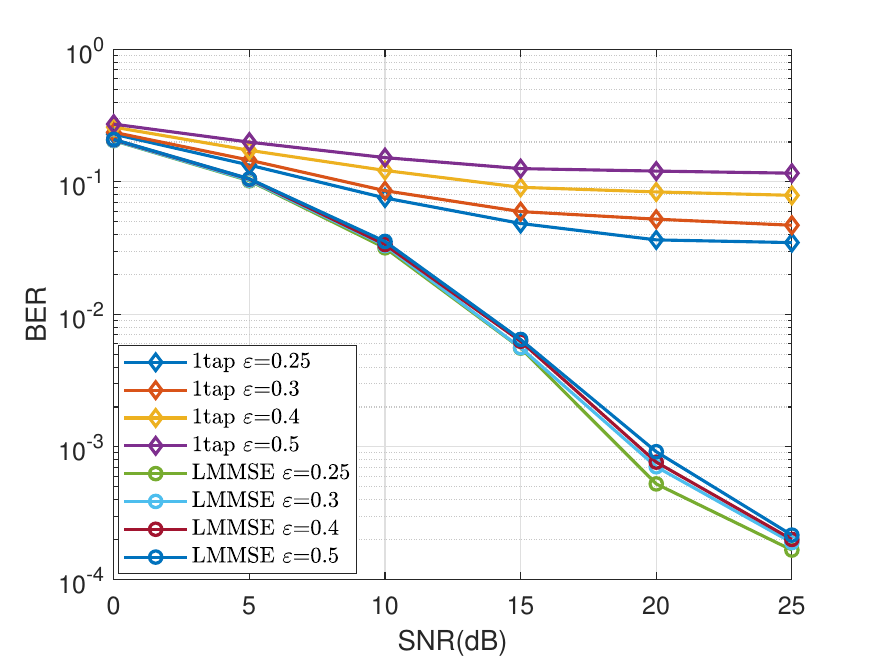}
		\label{BER of LMMSE and 1tap under NTN-TDL-B channel}
	}
	\quad
	\subfloat[BER performance of practical OFDM symbol detection with various CFO settings under the NTN-TDL-B channel model]
	{
		\includegraphics[width=0.9\linewidth]{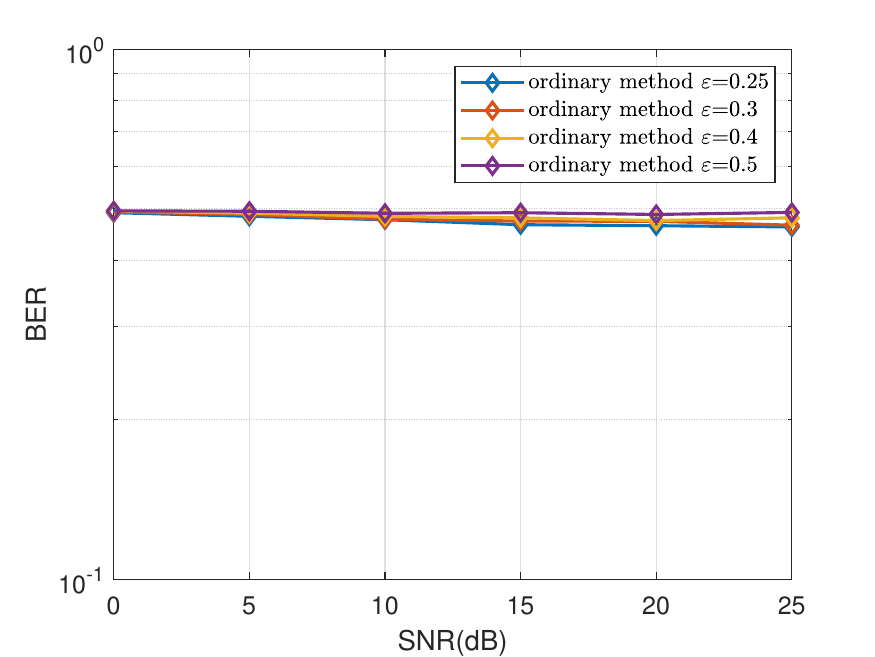}
		\label{BER of CFO compensation under NTN-TDL-B channel}
	}
	\caption{BER performance with various CFO settings under the NTN-TDL-B channel model}
	\label{BER NTN_TDL_B}
\end{figure}
\begin{figure}[htbp]
	\centering
	\subfloat[BER performance of practical OTFS symbol detection and ideal OFDM symbol detection with various CFO settings under NTN-TDL-D channel model]
	{
		\includegraphics[width=0.9\linewidth]{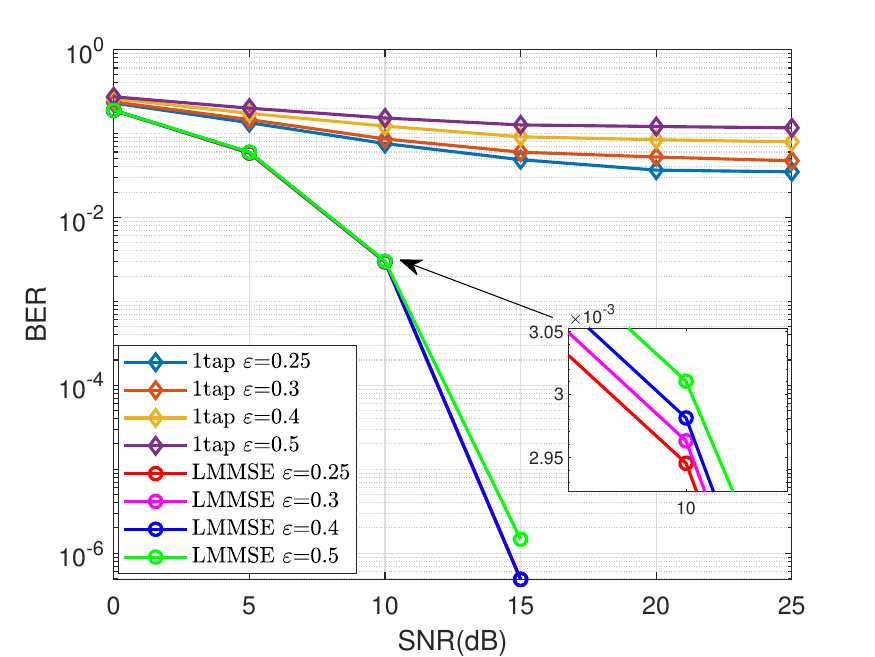}
		\label{BER of LMMSE and 1tap under NTN-TDL-D channel}
	}
	\quad
	\subfloat[BER performance of practical OFDM symbol detection with various CFO settings under NTN-TDL-D channel model]
	{
		\includegraphics[width=0.9\linewidth]{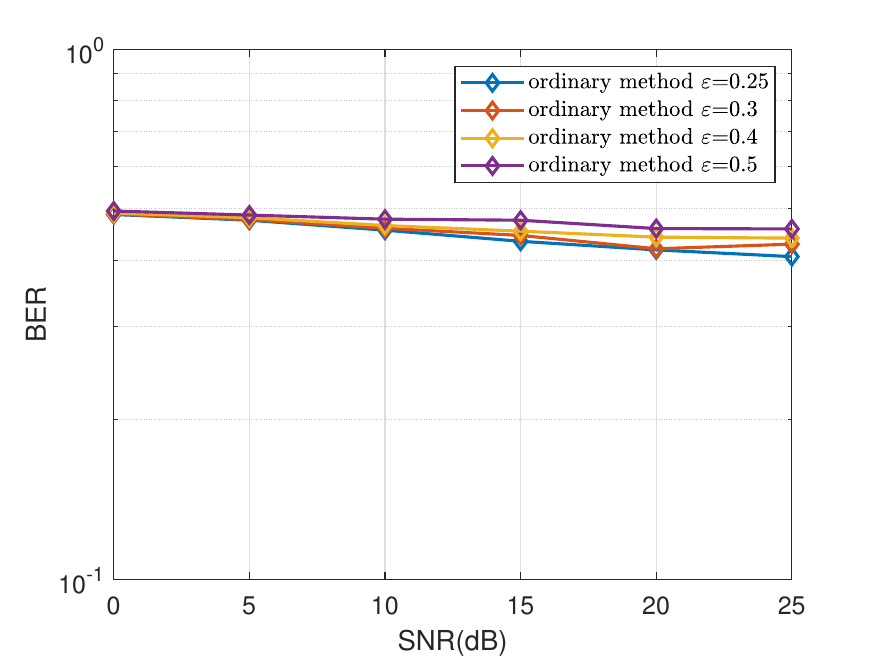}
		\label{BER of CFO compensation under NTN-TDL-D channel}
	}
	\caption{BER performance with various CFO settings under the NTN-TDL-B channel model}
	\label{BER NTN_TDL_D}
\end{figure}
It is observed in Fig. \ref{BER NTN_TDL_B}\subref{BER of LMMSE and 1tap under NTN-TDL-B channel}, the OTFS modulation has significantly superior performance to the OFDM modulation. Additionally, unlike the OFDM modulation, \rev{by which the BER tends to increase with the normalized CFO rising,} the OTFS modulation is revealed to be robust in the BER. This phenomenon reveals the robustness of the OTFS modulation in high-mobility scenarios with fast fading channels, and exhibits its capability to mitigate the phase ambiguity brought by CFO during symbols detection as well.    
Furthermore, as shown in Fig. \ref{BER NTN_TDL_D}\subref{BER of LMMSE and 1tap under NTN-TDL-D channel}, this performance advantage \rev{becomes} more pronounced in the NTN-TDL-D channel condition. This can be attributed to the structure of the NTN-TDL-D model, which comprises a dominant LOS path accompanied by several weaker multipath components, leading to relatively minimal MPSI and MUI. 
In practice, it is intractable to estimate channel state information for every block in the resource grid. Henceforth, unlike the LMMSE method for the OTFS modulation, the 1tap method is too ideal to be feasible in reality. 
In order to obtain the practical BER of the OFDM modulation, we employ Zadoff-Chu (ZC) sequence as pilots and estimate the channel state information through \rev{the} LS method. Meanwhile, we estimate the CFO via Moose method and detect the symbols with CFO compensation.   
In Fig. \ref{BER NTN_TDL_B}\subref{BER of CFO compensation under NTN-TDL-B channel} and Fig. \ref{BER NTN_TDL_D}\subref{BER of CFO compensation under NTN-TDL-D channel}, it exhibits the BER performance of the practical method is hardly tractable to detect symbols accurately in this scenario of high mobility. This is because that such a large CFO caused by high-speed movement of the LEO satellite reduces the accuracy of CFO estimation. Additionally, the disruption of orthogonality \rev{resulting} from the CFO leads to the ICI, which in turns causes extra phase ambiguity in symbols. Thus, the ordinary CFO compensation algorithm cannot mitigate the phase ambiguity for symbols in the high-mobility scenarios, like LEO satellite communication.

\subsection{Sum-rate with Proposed Algorithm}
\begin{figure}[htbp]
	\centering
	\subfloat[Comparison of the sum-rate performance between different OMA schemes and the proposed algorithm under the NTN-TDL-B channel.]{
		\includegraphics[width=0.9\linewidth]{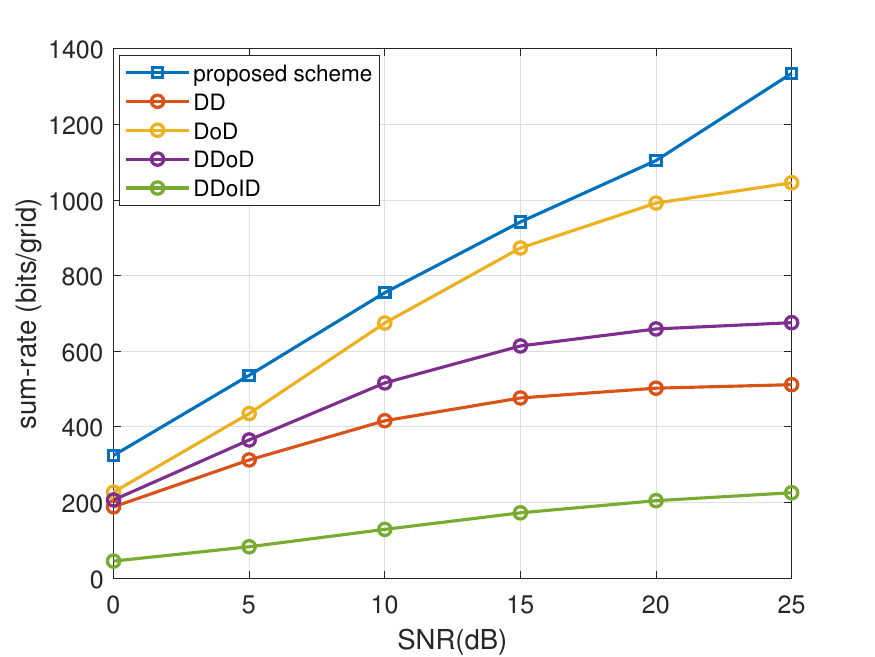}
		\label{system throughput of the proposed algorithm under NTN_TDL_B}
	}
	\quad
	\subfloat[Comparison of the sum-rate performance between different OMA schemes and the proposed algorithm under the NTN-TDL-D channel.]{
		\includegraphics[width=0.9\linewidth]{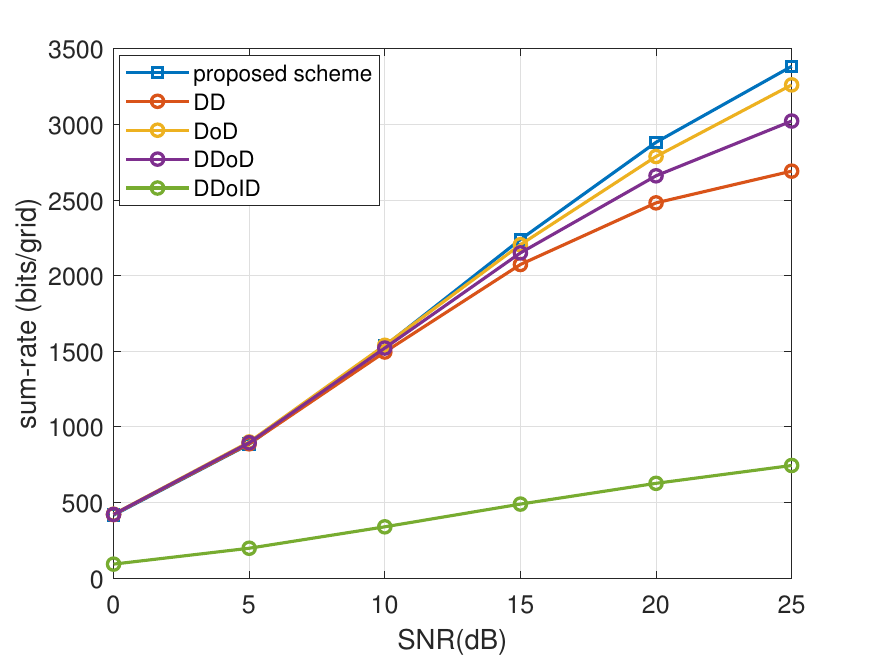}
		\label{system throughput of the proposed algorithm under NTN_TDL_D}
	}
	\caption{Comparison of the sum-rate performance between different OMA schemes and the proposed algorithm under diverse channel conditions.}
	\label{system throughput of the proposed algorithm}
\end{figure}

\begin{figure}[htbp]
	\centering
	\subfloat[The convergence curve of the proposed algorithm under the NTN-TDL-B channel.]{
		\includegraphics[width=0.9\linewidth]{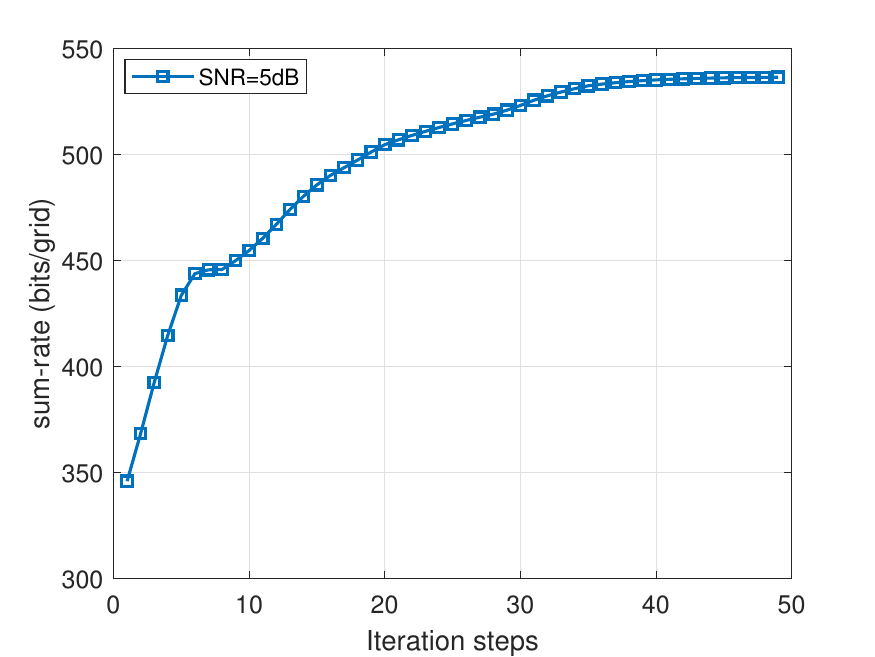}
		\label{convergence curve of the proposed algorithm under NTN-TDL-B channel}
	}
	\quad
	\subfloat[The convergence curve of the proposed algorithm under the NTN-TDL-D channel.]{
		\includegraphics[width=0.9\linewidth]{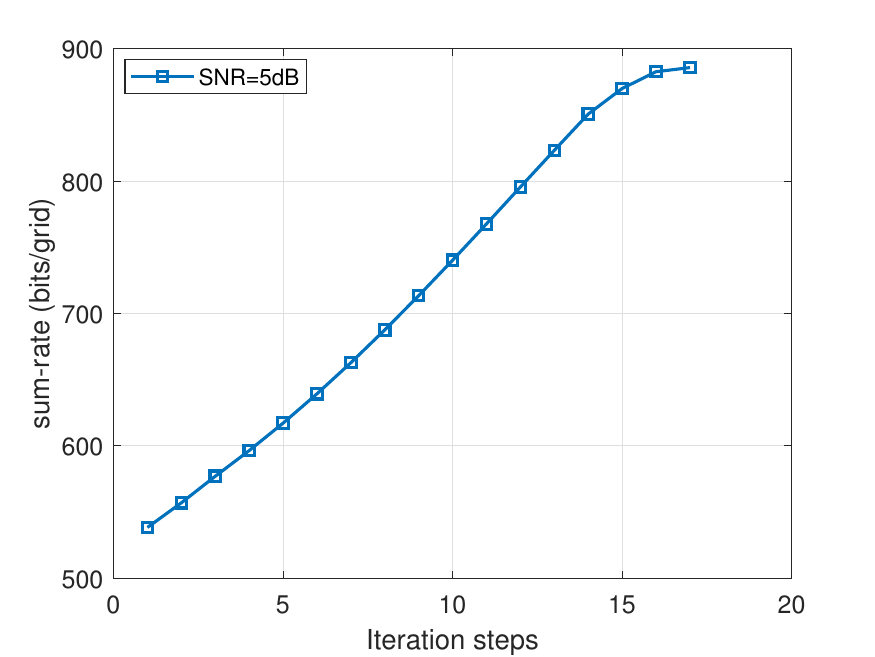}
		\label{convergence curve of the proposed algorithm under NTN-TDL-D channel}
	}
	\caption{The convergence curve of the proposed algorithm under diverse channel conditions.}
	\label{The convergence curve of the proposed algorithm}
\end{figure}

In this subsection, we present the sum-rate of the proposed penalty CCP algorithm with respect to different channel conditions. For both scenarios, we set $\varepsilon=0.25$. It can be seen from Fig. \ref{system throughput of the proposed algorithm}\subref{system throughput of the proposed algorithm under NTN_TDL_B}, the proposed algorithm outperforms all four ordinary OMA schemes in terms of the sum-rate. Furthermore, we also observe that the rate saturation phenomenon is mitigated at very high SNRs compared to the ordinary OMA schemes. This is because that the appropriate power allocation and symbols scheduling can mitigate the MPSI and MUI significantly, which makes the saturation slighter.
Moreover, as can be seen in Fig. \ref{The convergence curve of the proposed algorithm}\subref{convergence curve of the proposed algorithm under NTN-TDL-B channel}, the proposed algorithm converges in the finite steps. 


Furthermore, as seen from Fig. \ref{system throughput of the proposed algorithm}\subref{system throughput of the proposed algorithm under NTN_TDL_D}, the performance gain of the proposed algorithm in the NTN-TDL-D channel condition is larger than the gain in the NTN-TDL-D channel condition. This is because the NTN-TDL-D model is composed of a dominant LOS path and several much weaker multipaths, which results to relatively slight MPSI and MUI. Correspondingly, this characteristic reduces the computation complexity of the optimization procedure, as shown in Fig. \ref{The convergence curve of the proposed algorithm}\subref{convergence curve of the proposed algorithm under NTN-TDL-D channel}, the proposed algorithm can converge in less iterations. 

\section{Conclusion}
In this paper, we \rev{compared} the transmission performances of the OTFS modulation and the OFDM modulation in distinctive channel conditions with ordinary multiple access schemes, and \rev{proposed} a penalty CCP based algorithm for power allocation and symbols scheduling to replace the ordinary multiple access schemes. Furthermore, we \rev{investigated} performance and convergence of our proposed algorithm with respect to distinct channel conditions. Our simulation results \rev{showed} that the OTFS modulation is robust to the carrier frequency offsets and has superior performance to the OFDM modulation in relatively \rev{harsh} channel conditions. Moreover, the numerical results \rev{demonstrated} that our proposed algorithm outperforms the ordinary OTFS mutiple access schemes significantly.  

\begin{appendices}

\section{Orthogonal Multiple Access Schemes}
\begin{figure}[htbp]
	\centering
	\subfloat[DDMA]{
		\includegraphics[width=0.4\linewidth]{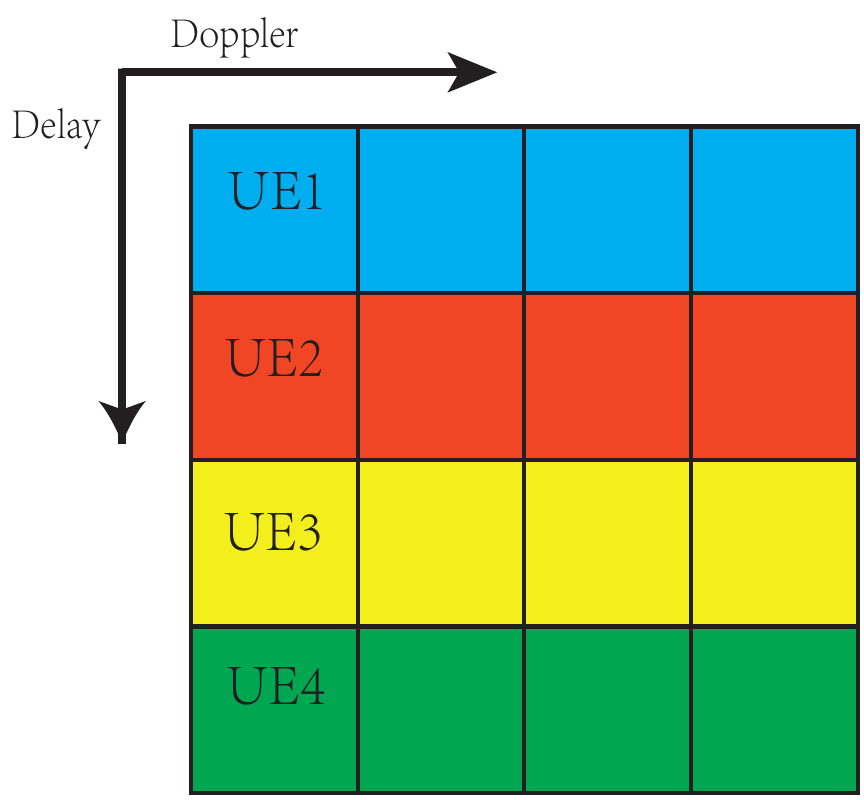}
		\label{DDMA}
	}
	\quad
	\subfloat[DoDMA]{
		\includegraphics[width=0.4\linewidth]{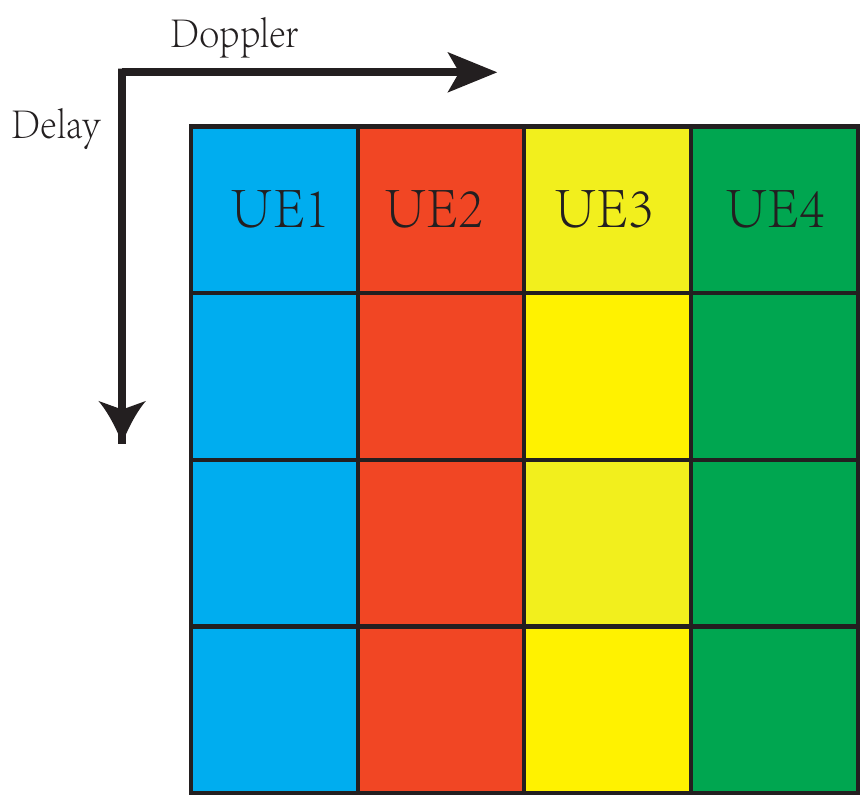}
		\label{DoDMA}
		}
	
	\subfloat[DDoDMA]{
		\includegraphics[width=0.4\linewidth]{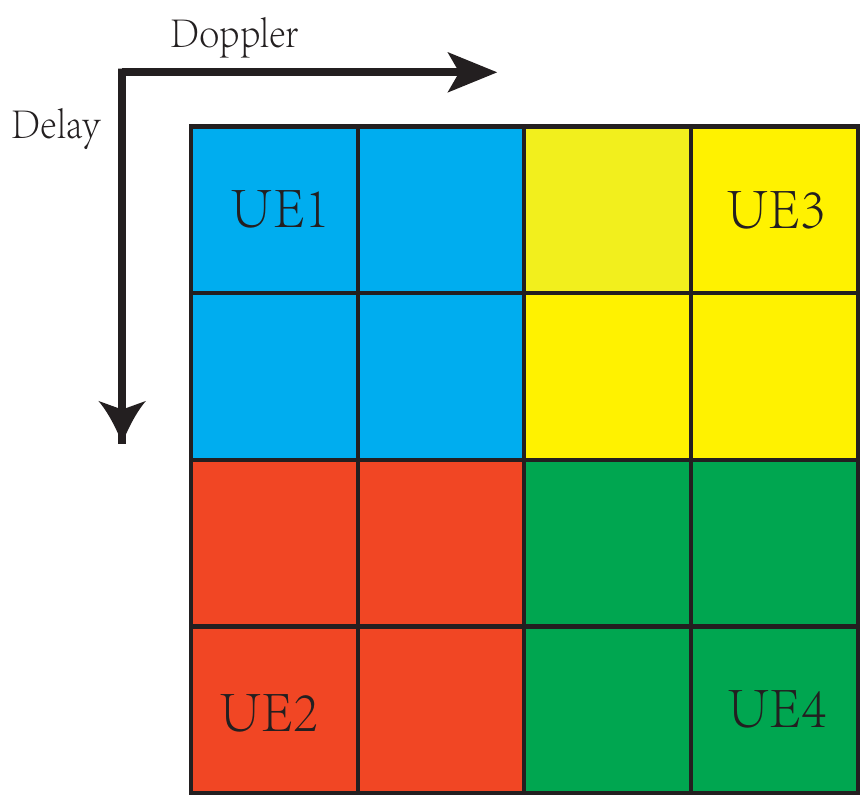}
		\label{DDoDMA}
		}
	\quad
	\subfloat[DDoIDMA]{
		\includegraphics[width=0.4\linewidth]{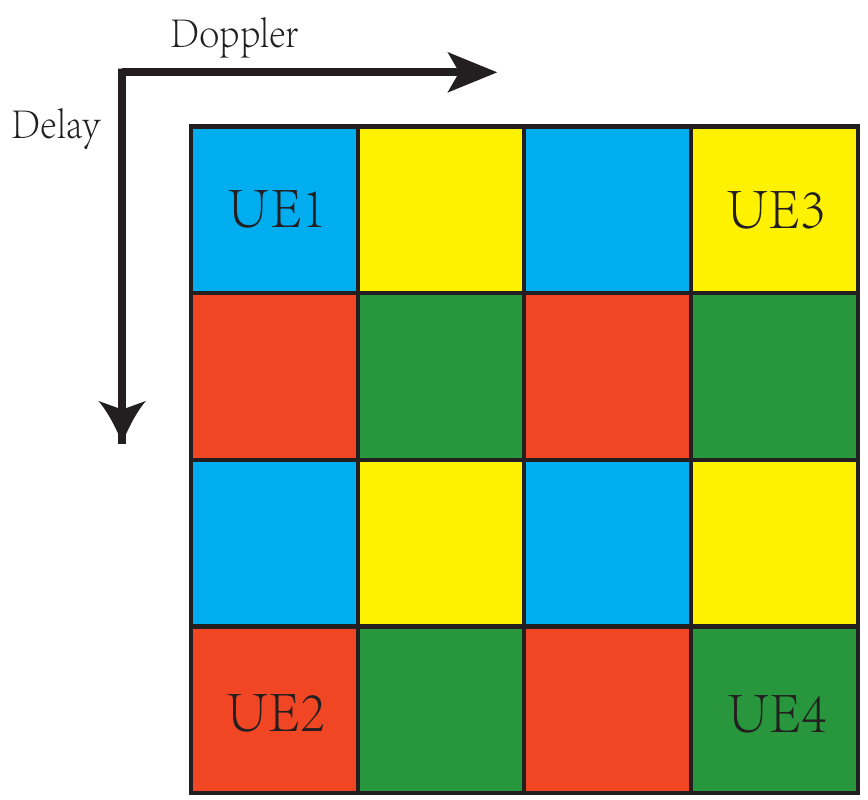}
		\label{DDoIDMA}}
	\caption{An Illustration of Different Multiple Access.}
	\label{An Illustration of Different Multiple Access}
\end{figure}

Fig. \ref{An Illustration of Different Multiple Access} presents several ordinary orthogonal multiple access schemes for allocating resource blocks to users in a downlink MU-SISO-OTFS system.
\subsubsection{DDMA (Multiplexing users along the delay axis)}
In this scheme, disjoint and contiguous bins along the delay axis are allocated to each user such that each user gets $M/K$ rows of the DD domain resource grid for transmission. The grid of the $i$-th user is defined as
\begin{equation}
	X_{\text{DD}}^{(i)}[l,k]=\begin{cases}
		a\in \mathbb{A} \quad &  \text{if} \; k\in\{0,1,\cdots,N-1\} , \\
		& \quad  l\in \left\{(i-1)\frac{M}{K},\cdots,i\frac{M}{K}-1\right\}, \\
		0 & \text{otherwise.}
	\end{cases}
\end{equation} 
Fig.~\ref{An Illustration of Different Multiple Access}\subref{DDMA} shows an example of this \rev{resource allocation scheme}, where an $M\times N=4\times 4$ grid in the DD domain \rev{is} allocated to four users.  Note that, although the users transmit on non-overlapping resource blocks, the symbols transmitted by each user experience MPSI and MUI as shown in \eqref{MU symbol-wise input out relation Expanded} due to the 2D circular convolution operation. 
\subsubsection{DoDMA (Multiplexing users along the Doppler axis)}
Similarly as the previous scheme, in this scheme non-overlapping and contiguous resource blocks along the Doppler axis are allocated to each user such that each user gets $N/K$ columns of the resource grid for transmission. The grid of the $i$-th user is defined as
\begin{equation}
	X_{\text{DD}}^{(i)}[l,k]=\begin{cases}
		a\in \mathbb{A} \quad & \text{if} \; k\in \left\{(i-1)\frac{N}{K}-1,\cdots,i\frac{N}{K}\right\}-1, \\
		& \quad l\in\{0,1,\cdots,M-1\}, \\
		0 & \text{otherwise.}
	\end{cases}
\end{equation} 
Fig.~\ref{An Illustration of Different Multiple Access}\subref{DoDMA} shows an example of this scheme allocation, where each user suffers from MUI and MPSI as well due to 2D circular convolution operation. 
\subsubsection{DDoDMA (Multiplexing Users along the delay-Doppler axis)}
In this scheme, non-overlapping and contiguous resource blocks along the delay axis and the Doppler axis are allocated to each user concurrently, thus each user gets a $N/\sqrt{K}\times M/\sqrt{K}$ block \rev{in the resource grid} of the $i$-th user is defined as
\begin{equation}
	X_{\text{DD}}^{(i)}[l,k]=\begin{cases}
		a \in \mathbb{A} & \text{if} \; k\in\left\{\Bigl\lfloor \frac{i}{\sqrt{K}} \Bigr\rfloor\frac{M}{\sqrt{K}},\cdots, \right.\\
		& \left. \qquad \quad \left(\Bigl\lfloor \frac{i}{\sqrt{K}} \Bigr\rfloor+1\right)\frac{M}{\sqrt{K}}-1\right\}, \\ 
		& \quad l\in\left\{\left([i-1]_{\sqrt{K}}\right)\frac{N}{\sqrt{K}},\cdots,\right. \\
		& \left. \qquad \quad \left([i]_{\sqrt{K}}\right)\frac{M}{\sqrt{K}}-1\right\}, \\
		0 & \text{otherwise.}
	\end{cases}
\end{equation}
Fig.~\ref{An Illustration of Different Multiple Access}\subref{DDoDMA} shows an example of this \rev{resource allocation scheme}, where each user suffers from MUI and MPSI as well due to 2D circular convolution operation. 
\subsubsection{DDoIDMA (Multiplexing Users as in \cite{2019OTFSMAMohammed})}
In previous schemes, each user's signal spans the entire TF domain. In \cite{2019OTFSMAMohammed}, the authors \rev{claim} that each user 's signal can be restricted to a sub-region of the TF domain. For $K=g_1g_2$ users, the allocation assumes that $g_1$ and $g_2$ divide $M$ and $N$ respectively. The $i$-th user ($0< i \le g_1g_2$) is allocated to
\begin{equation}
	X_{\text{DD}}^{(i)}[l,k]=\begin{cases}
		a\in \mathbb{A} \quad & \text{if} \; k=\bigl\lfloor (i-1)/g_1\bigr\rfloor+g_2u, \\
		& \quad l=[i-1]_{g_1}+g_1v, \\
		& \quad 0\le u<N/g_2, \; 0\le v<M/g_1, \\
		0, & \text{otherwise.}
	\end{cases}
\end{equation}
This scheme results in a periodic interleaving of symbols from each other as shown in Fig.~\ref{An Illustration of Different Multiple Access}\subref{DDoIDMA} for a system with $M=N=4$ and $g_1=g_2=2$. 

\section{Derivation of Achievable Rate for OFDM}
In this section, the symbol-wise input-output relation for the MU-SISO system is given based on the SU-SISO relation derived in \cite{hong2022delayDoppler}. We define the effective TF domain channel matrix for the $i$-th user as
\begin{equation}
	\mathbf{H}_{\text{TF}}^{(i)}=\left(\mathbf{I}_N\otimes \mathbf{F}_M\right)\mathbf{H}_{\text{TD}}^{(i)}\left(\mathbf{I}_N\otimes \mathbf{F}_M^\dagger\right).
\end{equation}
Correspondingly, the received information symbol vector $\mathbf{y}_{\text{TF}}^{(i)}$ in \eqref{y_TF} can be rewritten as 
\begin{equation}
	\mathbf{y}_{\text{TF}}^{(i)}
	\triangleq \mathbf{H}_{\text{TF}}^{(i)}\mathbf{x}_{\text{TF}}^{(i)}+\mathbf{w}_{\text{TF}}^{(i)},
\end{equation}
where $\mathbf{w}_{\text{TD}}^{(i)}$ the noise vector with the one side PSD $N_0$ for the $i$-th user within the TF domain.
For ease of derivation, we split the TF domain samples into $N$ blocks of length $M$ as
\begin{equation}
	\mathbf{x}_{\text{TF}}^{(i)}=\begin{bmatrix}
		\check{\mathbf{x}}_0^{(i)} \\ \vdots \\ \check{\mathbf{x}}^{(i)}_{N-1}
	\end{bmatrix}, \quad 
	\mathbf{y}_{\text{TF}}^{(i)}=\begin{bmatrix}
		\check{\mathbf{y}}_0^{(i)} \\ \vdots \\ \check{\mathbf{y}}^{(i)}_{N-1}
	\end{bmatrix},
\end{equation}
where $\check{\mathbf{x}}_n,\check{\mathbf{y}}_n \in \mathbb{A}^{M\times1}$.
Henceforth, the TF domain block-wise input-output relation can be defined as
\begin{small}
	\begin{equation}
		\begin{aligned}
			\check{\mathbf{y}}_0^{(i)}= & \sum_{j=1}^{K}\check{\mathbf{H}}_{0,0}^{(i)}\check{\mathbf{x}}_0^{(j)} + \check{\mathbf{w}}_0^{(i)}, \;  n=0, \\
			\check{\mathbf{y}}_n^{(i)}= & \sum_{j=1}^{K}\check{\mathbf{H}}_{n,0}^{(i)}\check{\mathbf{x}}_n^{(j)} + \sum_{j=1}^{K}\check{\mathbf{H}}_{n,1}^{(i)}\check{\mathbf{x}}_{n-1}^{(j)}+\check{\mathbf{w}}_n^{(i)}, \; 1\le n\le N-1 ,
		\end{aligned}
	\end{equation}
\end{small}
where $\check{\mathbf{H}}_{n,n'}^{(i)}\in\mathbb{C}^{M\times M}$ is the TF domain channel between the $n$-th received block and the $(n-n')$-th transmitted block for the $i$-th user with $n'\in \{0,1\}$, and $\check{\mathbf{w}}$ is the TF domain noise. Furthermore, the symbol-wise input-output relation in the TF domain can be rewritten as 
\begin{equation}
	\begin{aligned}
		\check{y}_0^{(i)}[m]=& \underbrace{\check{H}_{0,0}^{(i)}[m,m]\check{x}_0^{(i)}[m]}_{\text{Desired signal}} \\
		& + \underbrace{\sum_{j=1}^{K}\sum_{m'\neq m}\check{H}_{0,0}^{(i)}[m,m']\check{x}_0^{(j)}[m']}_{\text{ICI}} + \underbrace{\check{w}^{(i)}_0[m]}_{\text{noise}},\; n=0,
	\end{aligned}
\end{equation}
and
\begin{equation}
	\begin{aligned}
		\check{y}_n^{(i)}[m]=& \underbrace{\check{H}_{n,0}^{(i)}[m,m]\check{x}_n^{(i)}[m]}_{\text{Desired signal}} \\
		& + \underbrace{\sum_{j=1}^{K}\sum_{m'\neq m}\check{H}_{n,0}^{(i)}[m,m']\check{x}_n^{(j)}[m']}_{\text{ICI}} \\
		& + \underbrace{\sum_{j=1}^{K}\sum_{m'=0}^{M-1}\check{H}_{n,1}^{(i)}[m,m']\check{x}_{n-1}^{(j)}[m']}_{\text{ISI}} \\
		& + \underbrace{\check{w}^{(i)}_n[m]}_{\text{noise}}, \quad n=1\cdots N-1.
	\end{aligned}
\end{equation}
where the interference between the samples in the TF domain can be divided into ICI and inter-symbol interference (ISI) due to channel Doppler spread and delay spread, and AWGN $\check{w}$ has one-sided power spectral density $N_0$. 
 

Henceforth, the achievable rate $\check{R}_{m,n}^{(i)}$ can be given by 
\begin{equation}
	\check{R}_{m,0}^{(i)}=\frac{1}{2}\log_2\left(1+\frac{\bigl|\check{H}_{0,0}^{(i)}[m,m]\bigr|^2\check{\rho}_{m,0}^{(i)}}
	{\sum_{j=1}^{K}\sum_{m'\neq m}\bigl|\check{H}_{0,0}^{(i)}[m,m']\bigr|^2\check{\rho}_{m',0}+N_0}\right),
\end{equation}
for $n=0$ and
\begin{equation}
	\begin{aligned}
		\check{R}_{m,n}^{(i)}=&\frac{1}{2}\log_2\Biggl(1+\bigl|\check{H}_{n,0}^{(i)}[m,m]\bigr|^2\check{\rho}_{m,n}^{(i)} \bigg/ \\
		&\left(\sum_{j=1}^{K}\sum_{m'\neq m}\bigl|\check{H}_{n,0}^{(i)}[m,m']\bigr|^2\check{\rho}_{m',n}^{(i)} \right. \\
		& \; \left.\left. +\sum_{j=1}^{K}\sum_{m'=0}^{M-1}\bigl|\check{H}_{n,1}^{(i)}[m,m']\bigr|^2\check{\rho}_{m',n-1}^{(i)}+N_0\right)\right),
	\end{aligned}
\end{equation}
for $n=1,\cdots,N-1$,
where $\check{\rho}_{m,n}^{(i)}$ is the power allocated to the $(m,n)$-th symbol for the $i$-th user in the TF domain.

\end{appendices}

\bibliographystyle{IEEEtran}
\bibliography{myre}

\end{document}